\documentclass[
preprint,
showpacs,preprintnumbers,showkeys,
amsmath,amssymb,
aps,
pre,
]{revtex4-1}

\usepackage{graphicx}
\usepackage{hyperref}

\DeclareMathOperator\erf{erf}
\DeclareMathOperator\erfc{erfc}

\begin{document}

\author{G. R. Lee-Dadswell}
\email{geoffrey\_lee-dadswel@cbu.ca}
\affiliation{Math, Physics and Geology Department, Cape Breton University, Sydney, NS, Canada} 

\author{Nicholas Barrett}
\affiliation{Math, Physics and Geology Department, Cape Breton University, Sydney, NS, Canada}
\author{Michael Power}
\affiliation{Math, Physics and Geology Department, Cape Breton University, Sydney, NS, Canada}
%\author{Scott Cameron}  Include on next one
%\author{Kevin Moore}  Include on next one

\title{Cluster sizes in a classical Lennard-Jones chain}
\date{\today}

\begin{abstract}
The definitions of breaks and clusters in a one-dimensional chain in equilibrium are discussed.  Analytical expressions are obtained for the expected cluster length, $\langle K \rangle$, as a function of temperature and pressure in a one-dimensional Lennard-Jones chain.  These expressions are compared with results from molecular dynamics simulations.  It is found that $\langle K \rangle$ increases exponentially with $\beta = 1/k_BT$ and with pressure, $P$ in agreement with previous results in the literature.  A method is illustrated for using $\langle K \rangle (\beta, P)$ to generate a ``phase diagram'' for the Lennard-Jones chain.  Some implications for the study of heat transport in Lennard-Jones chains are discussed.
\end{abstract}



\maketitle

\section{Introduction}

% General discussion of one-dimensional chains
One-dimensional chains of classical particles have been an active area of research since the work of Fermi, Pasta, and Ulam over 60 years ago \cite{pap:Ford_review}.  As in that pioneering work, one-dimensional chains are often viewed as a ``toy model'' in attempts to understand phenomena in more complicated higher dimensional systems.  However, as the work by Fermi, Pasta, and Ulam clearly shows, one-dimensional chains can exhibit behaviours which are distinct from those of higher dimensional systems.  This has posed challenges to theorists.  A good example of this is transport.  Transport in one-dimensional systems is often anomalous \cite{book:Lepri_ThermalTransportLowDimensions}.  This has been seen for diffusion \cite{pap:hahn_etal_single-file_diffusion_observation,pap:metzler_klafter_random_walk_guide_anomalous_diffusion}, momentum transport \cite{pap:LNG_JStatPhys132} and heat transport \cite{book:Lepri_ThermalTransportLowDimensions,pap:Dhar2008,pap:chiu_etal_ballistic_phonon_thermal_transport_carbon_nanotube}.  As a result, classical and quantum mechanical one-dimensional systems have been a playground for theorists attempting to understand the underlying mechanical basis of constitutive relations such as Fourier's law of heat conduction \cite{pap:Bonetto_Lebowitz_FourierLaw}.  Many other examples of theoretical insights that arose from the study of one-dimensional chains could be given.  Perhaps one of the most important was the discovery of solitons, which resulted directly from attempts to understand the so-called ``FPU paradox'' \cite{pap:Ford_review,pap:Zabusky_Kruskal_PRL15}.

More practically, it has long been known that there are real physical systems which can be well modeled as one-dimensional classical chains.  For example, the scattering function for a classical chain was found to be in good agreement with neutron and X-ray scattering from the Hg chains in Hg$_{3-\delta}$AsF$_6$ \cite{pap:EmeryAxe_PRL40,pap:YoshidaShobuMori}.  The study of one-dimensional chains has led to a better understanding of dynamics and transport in a variety of other similar systems having subsystems which can be thought of as one-dimensional, such as zeolites \cite{Valiullin_Karger_inKlages_AnomalousMolecularDynamicsConfinedSpaces}.   More recently, our growing ability to fabricate and manipulate nanoscale structures such as nanotubes and nanowires \cite{pap:cahill_etal_nanoscale_thermal_transport,pap:Liu_etal_AnomalousHeatConductionAndAnomalousDiffusionInLowDimensionalNanoscaleSystems} and polymer chains \cite{pap:Meier_etal_LengthDependentTransportInPolymerChains_Experiment_PRL113} which are ``approximately one-dimensional'' has produced additional interest in one-dimensional transport.  These real systems are certainly not really one-dimensional in any sense.  However, the study of one-dimensional transport has yielded insights into the phenomenology of transport in such systems \cite{book:Lepri_ThermalTransportLowDimensions}.

The work presented in this paper is not directly about transport in one-dimensional systems.  However, it is motivated by the authors' interest in one-dimensional transport and is part of a broader study of transport in Lennard-Jones chains.  In particular, in later papers we wish to be able to understand the effect of chain fragmentation on the current power spectra of Lennard-Jones chains.  Hence, in motivating this work we will repeatedly refer to considerations of transport, even though transport is not our present topic.  This will be discussed further at the end of the paper.  For now we will make some brief comments about some outstanding problems in the study of one-dimensional transport to provide some motivation for examining the Lennard-Jones chain.

In one-dimensional chains which are known to exhibit anomalous thermal transport it is generally observed that the low frequency part of the heat current power spectrum goes as a simple power law $| j_\epsilon (\omega) |^2 \sim \omega^{-\alpha}$ \cite{pap:LNG_JStatPhys132,pap:Lepri_review}.  There is considerable controversy over the value of the exponent, $\alpha$.  However, it has begun to emerge that systems may fall into at least two universality classes, each with its own value of $\alpha$ \cite{pap:mine_universality_classes,book:Lepri_ThermalTransportLowDimensions,pap:Prahofer_Spohn_ExactScalingFunctionsForOne-DimensionalStationaryKPZ}.  While it is still a matter of debate what system properties determine which universality class a system falls into, it seems clear that whether the interparticle potential is symmetric plays a role.  Thus, the FPU-$\beta$ system (with a symmetric interparticle potential) is seen to possibly fall into a different class than the FPU-$\alpha \beta$ system.  However, the situation may not be quite so simple as this \cite{pap:mine_universality_classes}.  In any case, there is interest in examining chains with a wide variety of different interparticle potentials in order to assess which universality class they fall into.

A related issue is the ongoing effort to determine what the criteria are for a one-dimensional chain to exhibit normal heat transport.  Integrable systems, such as the harmonic chain and the hard point gas exhibit anomalous transport \cite{pap:Casati_Prosen_AnomalousConductionHardPointGas}.  Chains with smooth, unbounded potentials, such as FPU chains, are also known to exhibit anomalous transport \cite{book:Lepri_ThermalTransportLowDimensions}.  On the other hand, chains which violate momentum conservation, such as the ``ding-a-ling'' system \cite{pap:CasatiFordetal} and the Frenkel-Kontorova system \cite{pap:HuLiZhao_frenkel_kontorova} seem to have finite thermal conductivity.  However, only a few one-dimensional momentum conserving systems have been identified which exhibit normal transport.  These include the coupled-rotor system \cite{GiardinaLivietal} and, more controversially, the ``momentum conserving ding-a-ling'' \cite{pap:mine_mcdl}.

Transport has been previously examined in the Lennard-Jones chain.  Early results seem to be rather inconclusive about whether it has a finite thermal conductivity or not \cite{pap:bishop_collective_modes_LJ,pap:Mareschal_Amellal_PRA37}.  Some more recent investigators suggest that the Lennard-Jones chain may have a finite thermal conductivity \cite{pap:Li_etal_ConundrumOfAnomalousVsNormalHeatTransport,pap:Gendelman_Savin_NormalHeatConductivityInChainsCapableOfDissociation}.  This is certain to be a controversial claim and more work must be done on this system.  If it turns out that heat transport is anomalous in the Lennard-Jones chain then it will be of interest to determine what universality class it falls into.  This paper will not get as far as investigating these issues in any depth; it is preparatory to future papers which we hope will examine this.

In the present context, what is meant by a one-dimensional chain is a group of classical particles constrained to move along a line interacting with each other via some pair potential.  In many studies there are external non-conservative forces acting on the particles, such as a heat bath (e.g. \cite{pap:dunkel_etal_coherent_motions_dissipative_Morse,pap:pereira_anharmonic_crystals_as_thermal_diodes}).  Because we are working towards studying current power spectra of Hamiltonian chains we will not consider systems subject to non-conservative forces.  A chain with no non-conservative forces is described by the Hamiltonian

\begin{equation}
\mathcal{H}(\vec{x}, \vec{p}) = \sum_{i=1}^N \left [ \frac{{p_i}^2}{2m_i} + U (x_i) + \sum_{j \neq i} V (x_j - x_i) \right ],
\end{equation}

\noindent where $p_i$ is the momentum of the $i^{th}$ particle, $x_i$ is the position of the $i^{th}$ particle, $m_i$ is the inertia of the $i^{th}$ particle, $U$ is an ``on-site'' potential, perhaps modeling interactions of the system with some external agent, and $V$ is a pair potential modeling interactions between particles in the system.  There has been a great deal of interest in the role of the on-site potential in studies of one-dimensional transport (e.g. \cite{pap:HuLiZhao_frenkel_kontorova,pap:CasatiFordetal}) and particularly in the observation that the presence of on-site potentials may result in one-dimensional systems which obey Fourier's law.  However, we will not include on-site potentials in our system for two reasons.  Firstly, we are interested in the nature of the divergence of the current power spectra which occurs in systems with no on-site potentials.  Secondly, we are examining the role of chain fragmentation, and on-site potentials would tend to suppress this.

It is common to choose nearest-neighbour interactions for the pair potential.  This is sometimes justified by noting that interatomic forces in real solids have short range, and so it might be expected to be a good approximation \cite{pap:Bonetto_Lebowitz_FourierLaw}.  Additionally, the dominant mechanisms of transport in one-dimensional systems seem to be large scale collective motions \cite{pap:LNG_JStatPhys132,pap:spohn_nonlinear_fluct_hyd,pap:Delfini_etal} and so one might expect the microscopic details of the interparticle forces to play a small role.  In any case, the choice of nearest-neighbour interactions makes the analytical theory tractable and additionally makes it possible to carry out simulations much faster.  For all of these reasons we will choose to work with nearest-neighbour interactions.

It is useful to distinguish among several types of pair potentials: \emph{purely repulsive interactions} (such as the repulsive Coulomb interaction or the hard point gas), \emph{tightly bound chains} which we will define as chains in which as particle separation $r \rightarrow \infty$ the value of $V(r) \rightarrow \infty$ (such as the harmonic chain or the FPU-$\beta$ system), and \emph{loosely bound chains} which we will define as chains in which as $r \rightarrow \infty$ the potential goes to some finite value (such as the Lennard-Jones chain and the Morse chain).  Clustering and fragmentation is a characteristic behaviour of loosely bound chains.  We can now summarize our choice of systems.  We will examine one-dimensional Hamiltonian chains with no non-conservative forces and no one-site potentials.  Some of our results will be applicable to all such systems which can be categorized as loosely bound chains, but the bulk of our results will be specific to the one-dimensional Lennard-Jones chain.

While exceptions are known (e.g. \cite{pap:dyson_existence_of_phase_transition_one-dimension,pap:WangCasati_1DPhaseTransition}), most one-dimensional systems do not exhibit true phase transitions \cite{pap:van_hove_physica_no_one-dimensional_phase_transition,LandauLifshitz_StatPhys}.  It may be possible to distinguish among rather distinct states of a one-dimensional system.  However, as the thermodynamic parameters of the system such as temperature and pressure are changed there are no discontinuous changes in any system properties.  That is, the transitions from one state to another are smooth and so a continuum of intermediate states exist.  More concretely, for one-dimensional, loosely bound chains at sufficiently low temperature and moderate pressure the system tends to exist in a state where all particles are separated by a distance close to the equilibrium bond length.  This has been referred to as a ``solid'' state \cite{pap:chetverikov_dunkel_phase_behaviour_Morse}.  However, given the lack of long-range order in one-dimensional chains \cite{pap:YoshidaShobuMori} it might be more accurate to call it a ``liquid'' state.  At sufficiently high temperature and the same pressure the system must eventually be in a state where all particles are unbound from their neighbours and typical interparticle distances are much larger than the equilibrium distance.  It seems quite reasonable to call this a ``gas'' state of the system.  However, in going from the liquid state to the gas state there is no well defined phase transition where the volume as a function of temperature is discontinuous.  Instead the system is seen to break up into clusters and as the temperature is increased the average cluster size becomes smaller in a continuous fashion.  The only exception to this is that at zero temperature and sufficiently high density the system has a genuinely crystaline structure.  This ordered state is destroyed discontinuously in moving from zero temperature to any arbitrarily small non-zero temperature \cite{pap:stillinger_stat_mech_of_metastable_matter,pap:Lepri_etal_1d_LennardJones}.  We will restrict our attention to the non-zero temperature case where no true phase transitions exist.

Despite the lack of well defined phases of loosely bound one-dimensional chains there has been some interest in studying the ``phase behaviour'' of Lennard-Jones and Morse chains.  Some of the interest in this topic \cite{pap:chetverikov_dunkel_phase_behaviour_Morse,pap:chetverikov_etal_thermo_phase_trans_Morse} stems from clustering in these chains being an easily studied example of a collective thermal excitation.  Other work in this area \cite{pap:Lepri_etal_1d_LennardJones} is connected with a desire to understand long-time behaviour of various hydrodynamic correlation functions.  Ultimately, both of these are related to transport in the system and so the interest in the phase behaviour of these systems can been seen as related to the study of one-dimensional transport.

The work on phase behaviour of chains is also somewhat related to the study of fragmentation or fracture of materials.  In this paper we will have reason to refer to some results from the study of fracture in one-dimensional chains and so it is worth briefly reviewing it.  The interest in fracture of one-dimensional chains stems in part from the need to understand polymer degradation and scission of DNA and other biological macromolecules \cite{pap:oliveira_trans-state_frac_nuc}.  Like one-dimensional transport, fracture in one-dimension is somewhat anomalous.  In particular, unlike higher dimensional systems, a one-dimensional loosely bound chain will fracture (eventually) if is subjected to any arbitrarily small externally applied tension.  This was conjectured in \cite{pap:welland_etal_fracture_in_1d}.  A reasonable ``proof'' that this must be true is to simply observe that for loosely bound chains, the equilibrium length per particle as a function of pressure diverges to infinity as the pressure $P \rightarrow 0$ and is infinity for any negative pressure.  So a finite length chain under tension is in a non-equilibrium state.

In the discussion that follows it will be useful to distinguish among several types of studies of fracture and clustering in loosely bound chains.

\begin{itemize}

\item \textbf{Non-equilibrium chain fracture studies}: An unbroken chain is put under tension and monitored until it breaks.  Past research studies have included examinations of the criteria for bond stability \cite{pap:oliveira_bond_stability_criterion}, how breaking time depends on temperature, stress and other factors \cite{pap:oliveira_trans-state_frac_nuc}, and the nature of the redistribution of the strain field in the chain after fracture occurs \cite{pap:sain_etal_rupture_extended_object}.

\item \textbf{Active chain phase behaviour}: A chain in periodic boundary conditions is placed in contact with a heat bath and the degree of clustering is examined as a function of heat bath parameters, chain density, etc. \cite{pap:dunkel_etal_coherent_motions_dissipative_Morse,pap:chetverikov_dunkel_phase_behaviour_Morse,pap:chetverikov_etal_thermo_phase_trans_Morse,pap:sain_etal_rupture_extended_object}

\item \textbf{Passive chain phase behaviour}: A chain in periodic boundary conditions is initialized into an equilibrium state and the degree of clustering is examined as a function of chain energy and density while the chain evolves subject only to its own internal interactions \cite{pap:chetverikov_dunkel_phase_behaviour_Morse,pap:chetverikov_etal_thermo_phase_trans_Morse,pap:Lepri_etal_1d_LennardJones}.

\end{itemize}

\noindent Our specific interests in transport theory direct our attention to Hamiltonian chains, and so we will restrict our attention to passive chains.  However, as will be explained below, some of our choices with regards to defining breaks in our chains are motivated by findings in studies on active chains which is why we have included the previous descriptions.

A topic which was briefly examined in \cite{pap:Lepri_etal_1d_LennardJones} was how the fraction of broken bonds in a Lennard-Jones chain in equilibrium depends on temperature.  Completely equivalently, one could talk about how the expected cluster length depends on temperature.  This was examined numerically in \cite{pap:Lepri_etal_1d_LennardJones}.  The work in the present paper complements the earlier numerical work.  We will develop an analytical theory which predicts the expected cluster length in a Lennard-Jones chain as a function of temperature and pressure.  We will then illustrate how this theory can be used to generate a phase diagram for the chain similar to the one obtained from simulations in \cite{pap:chetverikov_dunkel_phase_behaviour_Morse}.  The theory is developed from nothing more complicated than equilibrium statistical mechanics.

Additionally, thermodynamic properties for the Lennard-Jones chain such as energy-density-temperature-pressure relations \cite{pap:bazhenov_heyes_dynprops_transcoeffs_LJ}, specific heat capacity \cite{pap:Mareschal_Amellal_PRA37}, and the thermodynamic speed of sound \cite{pap:Lepri_etal_1d_LennardJones} have previously been published in the literature.  These have generally been determined from simulations.  An exception is \cite{pap:kim_tadmor_analytical_free_energy_1d_chain} which presents a useful, fully analytical free energy expression for the Lennard-Jones chain.  However, this is done in an approximation which is only valid at low temperatures, though they compare their results with exact ones.  Surprisingly though, detailed derivations of exact expressions for these quantities have not been presented in the literature.  Perhaps such a derivation is considered ``trivial'' because it involves nothing more than standard equilibrium statistical physics and because in the end it must involve some integrals which have to be evaluated numerically.  Nevertheless, because of the continued interest in this system we feel that it would be useful to develop some exact expressions for the system's thermodynamic quantities which can be (numerically) evaluated reasonably easily.  We present these in this paper in the hopes that it will prove to be a useful resource.

\section{Clusters and Breaks in Loosely Bound Chains}

% Rewrite this whole section

In the theory development part of the paper we will be interested in the expected distance between ``breaks'' in a loosely bound chain, or equivalently, the expected cluster length.  We must begin by defining what we mean by a ``break'' in a chain, or equivalently, how we determine that some subset of the particles in the chain are all in the same cluster.  Broadly speaking, a break is a bond which has been stretched to a sufficiently long length that the interaction between the particles connected by the bond is ``very weak'' in some sense.  A cluster is a contiguous group of particles with breaks at both ends but no breaks in the middle.  However, beyond these rough definitions, a brief review of how these concepts have been defined by previous researchers shows that there is no established way to make these definitions precise; various researchers have used different definitions and some have explicitly pointed out the arbitrariness of their definitions.  Nevertheless, we will need to choose definitions that are suitable to our purposes.  We will be studying passive chains, so we could follow the definitions used in previous studies of passive chains.  However, there are insights to be gained from non-equilibrium chain fracture studies and these will influence our choice of definitions. 

It is worth reviewing some key facts about the conditions under which loosely bound chains break.  The only way for a chain to remain unbroken indefinitely is for it to be at zero temperature.  Even at zero temperature, if the tension in the chain is above some critical value then the lowest energy state of the chain is a broken state \cite{pap:stillinger_stat_mech_of_metastable_matter}.  At non-zero temperature, any finite-length loosely bound chain cannot be in equilibium when it is under tension.  So, for a chain pulled from its ends at non-zero temperature, if the chain starts unbroken it will always break given time \cite{pap:welland_etal_fracture_in_1d,pap:oliveira_taylor_breaking_polymer_chains}.  Under periodic boundary conditions or fixed boundary conditions a chain at nonzero temperature will repeatedly break and ``heal'', even if the chain is under pressure, so that the number of breaks in the chain will fluctuate with time.

The majority of the discussion of how to define chain breaks is found in the literature on non-equilibrium chain fracture.  Most simply one would define a \emph{breaking length} beyond which any bond is said to be broken.  For a loosely bound chain with smooth potential there is a critical bond length, $q_c$, at which the interparticle force is a maximum.  Naively one might expect that a chain pulled from its ends will break if any bond exceeds $q_c$.  Slightly less naively, for a fixed length chain one can define an effective potential for a bond \cite{pap:oliveira_bond_stability_criterion}.  For sufficient tension this effective potential always has a maximum at some length somewhat greater than $q_c$ and one would expect that the chain will break if a bond exceeds this length.  However, it turns out that both of these expectations are incorrect \cite{pap:oliveira_taylor_breaking_polymer_chains, pap:oliveira_bond_stability_criterion}.  Bonds can exceed these lengths by a surprising amount without the chain irreversibly breaking.  This is sometimes called ``bond healing''.  It results from the collective nature of the chain motions.  Simply put, if an individual bond is stretched there is a high probability that surrounding bonds are compressed and so the two particles connected by the stretched bond will tend to be pushed back together.  Thus, the fracture of a chain is an inherently nonlocal phenomenon.  For our purposes, the key insight from all of this is that $q_c$ seems not to be a reasonable choice for a breaking length.  Given that even a chain under tension can heal even when a bond exceeds $q_c$ by a large amount, it is reasonable to choose a longer breaking length than this.

In non-equilibrium chain fracture studies, where the chain breaks irreversibly it is possible to investigate what the breaking length is.  In contrast, in both active and passive chain studies the chain is generally studied in periodic boundary conditions.  Every break in the chain will eventually heal given time in this case.  The definition of a breaking length is more arbitrary as a result.  For example, in a study of active chains \cite{pap:chetverikov_dunkel_phase_behaviour_Morse} a critical density, $n_c$, is defined which corresponds to particles equally spaced by a distance equal to the bond length at which the maximum attractive force occurs.  A broken bond was then defined as a bond for which the particle separation $q > 1/n_c$.  This corresponds to the critical bond length, $q_c$, given above.  On the other hand, in a study of passive chains \cite{pap:Lepri_etal_1d_LennardJones} the authors chose a value for the critical particle separation that was $1.5/n_c = 1.5 q_c$.  However, the details of this choice are not likely to matter in the following sense.  Unlike in chain fracture studies, where it is important to be able to define an instant in time when the chain breaks, in these studies one is more concerned with various averages such as the average number of breaks in the chain.  The exact values of these averages will depend on the precise definition of a broken bond, but the qualitative behaviours of these averages as functions of temperature, pressure, and other external influences are not likely to depend on the definition.  This was commented upon and checked in \cite{pap:chetverikov_dunkel_phase_behaviour_Morse}.  However, recall the insights from chain fracture studies which show that it is reasonable to define a breaking length significantly longer than $q_c$.  Given this, when we give a purely geometric definition of broken bonds using only bond length, we will follow \cite{pap:Lepri_etal_1d_LennardJones} and use a longer breaking length, $1.5 q_c$.

Finally, it is worthwhile to mention one more insight from an active chain study which will motivate some of how we define broken bonds.  Consider a loosely bound chain in which $lim_{r \rightarrow \infty} V(r) = 0$;  this is a common choice for definitions of both the Lennard-Jones and Morse potentials.  In considering the phase behaviour of such a chain it may be useful \cite{pap:chetverikov_dunkel_phase_behaviour_Morse} to distinguish between clusters, bounded by breaks and with a total cluster energy $E_{cl} < 0$, as opposed to ``strong compressions'' also bounded by breaks but with $E_{cl} > 0$.  If the system is dominated by clusters in the first sense then it can be thought of as being in a liquid-like state that has broken up into ``droplets''.  If it is dominated by strong compressions then it is more like a high density gas with large density fluctuations.  Another way of thinking of this is that the dynamics of the chain are likely to different if neighbouring particles are likely to be bound ($E_{pair} < 0$) as opposed to the situation in which neighbouring particles are likely to be unbound ($E_{pair} > 0$) independently of the likelihood of a bond length to be above some breakage length.

\section{Thermodynamics of the Lennard-Jones Chain}
\label{sec:LJthermo}

In this section we present detailed expressions for key thermodynamic quantities of the Lennard-Jones chain.  A general formalism for determining all of this was presented in \cite{pap:spohn_nonlinear_fluct_hyd}.  The formalism presented there is far more compact and elegant than what is done here.  However, in the interest of having expressions specific to the Lennard-Jones chain which can be easily evaluated we have chosen to follow a process and notation more similar to the one used in \cite{pap:LNG_JStatPhys132} for FPU chains.

A one-dimensional chain consists of $N$ particles constrained to move along a line so that the state of the system is described by the particle positions, $\{x_1, \, x_2, \, \ldots ,\, x_N \}$, and particle momentums, $\{ p_1, \, p_2, \, \ldots , \, p_N  \}$.  The distances, $q_i \equiv x_{i+1} - x_i$, are ``bond lengths''.  As is often done, we will restrict our attention to the case of nearest neighbour interactions.  So the dynamics of the system depends only on $N$ bond lengths (rather than $N$ particle positions) and $N$ momenta.  In particular this means that the Hamiltonian only involves $N$ interparticle potentials.  Throughout this paper the interactions are governed by a Lennard-Jones interparticle potential.  The Hamiltonian of the system is

\begin{equation}
\mathcal{H}(\vec{q}, \vec{p}) = \sum_{i=1}^N \left [ \frac{{p_i}^2}{2m} + V_{LJ} (q_i) \right ],
\end{equation}

\noindent where $m$ is the particle mass.  The interparticle potential, $V_{LJ}$, is described by

\begin{equation}
V_{LJ} (q) = A \left [ \left ( \frac{q_0}{q} \right )^{12} - \left ( \frac{q_0}{q} \right )^6 \right ] .
\end{equation}

\noindent The parameter $A$ simply characterizes the overall strength of the interaction and is related to the ``well depth'' of the potential, $D$, via $D = A/4$.  The parameter, $q_0$, is the interparticle separation at which $V_{LJ} = 0$.  The equilibrium bond length is 

\begin{equation}
q_{min} = 2^{1/6} q_0 \, ,
\end{equation} 

\noindent and the critical bond length at which the maximum attractive force occurs is 

\begin{equation}
\label{eq:qc}
q_c = \left (\frac{13}{7} \right )^{1/6} q_{min} = \left (\frac{26}{7} \right )^{1/6} q_0 \, .
\end{equation}

The system is autonomous (momentum and energy conserving) and throughout the paper we work with periodic boundary conditions.  In practice this is done by defining an $(N+1)^{th}$ particle and fixing $q_{N+1} = q_1$, $p_{N+1} = p_1$.

We consider an isoenthalpic-isobaric ensemble (constant pressure canonical ensemble) of Lennard-Jones chains.  The partition function is

\begin{equation}
Y(\beta, P, N) = \frac{1}{h^N} \int d\vec{q} d\vec{p} \exp{[-\beta(\mathcal{H}(\vec{q}, \vec{p}) + PL(\vec{q}))]} \, ,
\end{equation}

\noindent where the prefactor involving Planck's constant is unimportant for our purposes, the integral is over the whole $2N$-dimensional phase space, $\beta = 1/k_B T$ is the inverse temperature, $P$ is the ensemble pressure, and $L = \sum_i q_i$ is the chain length.  Because of the choice to work with nearest-neighbour interactions, this splits into $N$ statistically independent factors

\begin{equation}
\label{eq:part_func}
Y(\beta, P, N) = \frac{1}{h^N} \left [ \int_{-\infty}^\infty dp \int_0^\infty dq F(q, p, \beta, P) \right ]^N \, ,
\end{equation}

\noindent with

\begin{equation}
F(q,p, \beta, P) = \exp{\left \{-\beta \left [ \frac{p^2}{2m} + A \left ( \left ( \frac{q_0}{q} \right )^{12} - \left ( \frac{q_0}{q} \right )^6 \right ) + Pq \right ]  \right \} } \, .
\end{equation}

We first note that if we work with some set of units characterized by length and time scales

\begin{eqnarray}
\ell_0 & = & (\beta A)^{1/6} q_0 \, , \\
t_0 & = & \left ( \frac{A}{2} \beta^4 m^3  \right )^{1/6} q_0 \, ,
\end{eqnarray}

\noindent so that we may define dimensionless variables, $\overline{q}$ and $\overline{p}$, by

\begin{eqnarray}
q & = & \ell_0 \overline{q} \, , \\
p & = & \frac{m\ell_0}{t_0} \overline{p} \,  
\end{eqnarray}

\noindent then the partition function reduces to

\begin{equation}
Y = [C(A / \beta)]^N \int_0^\infty d\overline{q} \exp{\left [ -\frac{A^\ast}{\overline{q}^{12}} + \frac{1}{\overline{q}^6} - P^\ast \overline{q} \right ]} \, ,
\end{equation}

\noindent where $C(A/ \beta)$ is a prefactor which depends only on the ratio, $A/ \beta$, and

\begin{eqnarray}
A^\ast & = & \frac{1}{\beta A} \, , \\
P^\ast & = & A^{1/6} \beta^{7/6} P q_0 \, ,
\end{eqnarray}

\noindent are dimensionless parameters.  This shows that for anything which is determined entirely by the statistics of the Boltzmann factors (such as the probability distribution of bond lengths) we can explore the entire relevant parameter space of the system by varying only two quantities, $A^\ast$ and $P^\ast$.  So, for example, varying the potential parameters, $A$ and $q_0$, is equivalent to varying $\beta$ and $P$.  Thermodynamic quantities which are related to the partition function via derivatives with respect to $\beta$ will additionally depend in a simple way on the ratio $A/\beta$.  However, the dependence on $A/\beta$ takes the form of a simple scaling and we need only vary two parameters to explore all of the possible thermodynamic regimes of the system.  We choose to vary $\beta$ and $P$ and fix $A = 1$, $q_0 = 1$.

Returning to (\ref{eq:part_func}), we can obtain all of the equilibrium thermodynamics by standard methods.  It is convenient to rewrite (\ref{eq:part_func}) as

\begin{equation}
Y(\beta, P, N) = \left [Z_p (\beta) \right ]^N \left [\Phi_0 (\beta, P) \right ]^N \, ,
\end{equation}

\noindent where

\begin{equation}
Z_p (\beta) = \sqrt{\frac{2 \pi m}{\beta}}
\end{equation}

\noindent is the usual momentum part of the partition function and where we define

\begin{equation}
\Phi_n (\beta, P) = \int_0^\infty dq \, q^n \exp{\left \{ -\beta \left [ V_{LJ} (q) + Pq \right ]  \right \} } \, .
\end{equation}

\noindent The set of functions, $\Phi_n$, are convenient to define because $\Phi_0$ appears in the partition function and $\Phi_n$ for various values of the index, $n$, will appear in the thermodynamic quantities which are obtained from the partition function.  In particular, calculations can be considerably simplified by noting the derivative identities

\begin{eqnarray}
\label{eq:bderiv}
\frac{\partial}{\partial \beta} \Phi_n & = & -A{q_0}^{12} \Phi_{n-12} + A {q_0}^6 \Phi_{n-6} - P \Phi_{n+1} \, , \\
\label{eq:Pderiv}
\frac{\partial}{\partial P} \Phi_n & = & -\beta \Phi_{n+1} \, .
\end{eqnarray}

\noindent Evaluation of the integrals contained in the $\Phi_n$ functions themselves can only be carried out numerically.  However, this does allow us at least to write expressions for the thermodynamic constants in fairly compact forms.  It is now quite straightforward to use the derivative identities (\ref{eq:bderiv}) and (\ref{eq:Pderiv}) to obtain all of the equilibrium thermodynamic quantities for the system.

\begin{eqnarray}
\label{eq:GLE}
G (\beta, P, N) & = & -\frac{1}{\beta} \ln{\left [ Y(\beta, P, N) \right ]} = -\frac{N}{\beta} \left [ \frac{1}{2} \ln{\left ( \frac{2\pi m}{\beta} \right )} + \ln{(\Phi_0)} \right ] \, ,\\
\label{eq:L}
\langle L \rangle & = & \left ( \frac{\partial G}{\partial P} \right )_\beta = N\frac{\Phi_1}{\Phi_0} \, ,\\
\label{eq:E}
\langle E \rangle & = & \frac{\partial}{\partial \beta} (\beta G) - P\langle L \rangle = \frac{N}{2\beta \Phi_0} \left [ 2A{q_0}^{12} \beta \Phi_{-12} -2A{q_0}^6 \Phi_{-6} + \Phi_0  \right ] \, .
\end{eqnarray}

\noindent where $G$ is the Gibbs free energy, $L$ is the system length and $E$ is the total system energy.  We may now calculate other equilibrium thermodynamic quanties from the usual derivatives: specific heat capacity $c_P = (-k_B \beta^2/N)\partial \langle (E+PL) \rangle /\partial \beta $, isothermal compressibility $\chi_T = -(1/\langle L \rangle) \partial \langle L \rangle /\partial P$, and thermal expansion coefficient $\alpha_P = -1/(k_BT^2 \langle L \rangle) \partial \langle L \rangle / \partial \beta$.  The expressions for these are too long to be easily printed here but they are easily obtained from (\ref{eq:L}) and (\ref{eq:E}) using the identities, (\ref{eq:bderiv}) and (\ref{eq:Pderiv}).  Additionally the thermodynamic speed of sound can be obtained from the well-known thermodynamic identity

\begin{equation}
c^2 = \frac{\ell \gamma}{m \chi_T}
\end{equation}

\noindent where $\ell = L/N$ is the average length per particle, $\gamma = c_P/c_V$ is the specific heat ratio, and we may obtain the constant volume specific heat from $c_V = c_P -\ell T {\alpha_P}^2/\chi_T $.

\begin{figure}
\includegraphics[width=\textwidth]{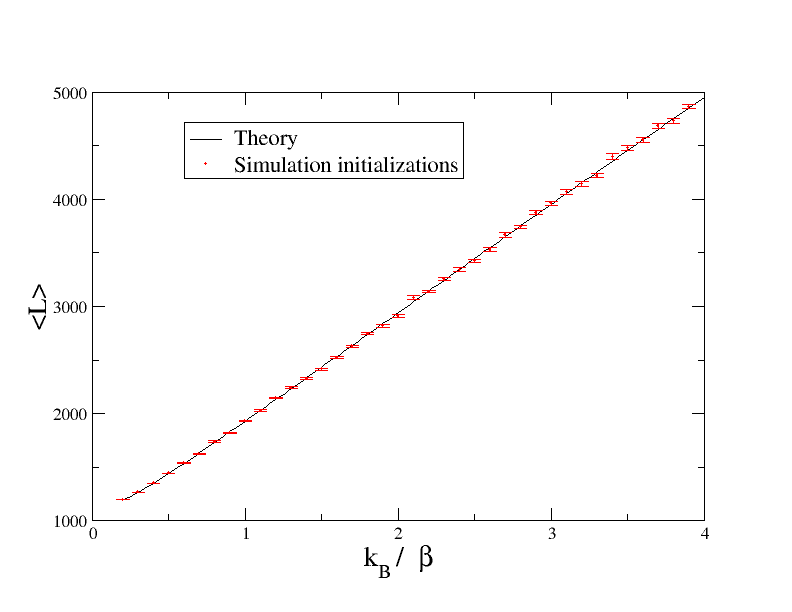}
\caption{\label{fig:LvsT}Expected chain length, $\langle L \rangle$, as a function of temperature for $P = 1$, $N=1024$ using both (\ref{eq:GLE}) and the initialization routines from simulations.  Error bars indicate one standard error in the results from 100 initializations for each $(P, T)$ of the simulations.}
\end{figure}

A major reason for obtaining the above expressions is that we may use them to validate the initialization routines used in our simulations.  An example of this validation is shown in Figure~\ref{fig:LvsT}.  The initialization protocol for the simulations will be discussed below.  Additionally in the conclusions we will make some use of the ability to find the speed of sound in the chain.

\section{Clusters Defined by Interparticle Distance}

The most straightforward way to predict cluster sizes is to define a ``break'' in the chain as any location where $q_i > \ell_{b}$ given some chosen ``breakage length'' $\ell_{b}$.  As already pointed out, any definition of when and where the chain is broken will be rather arbitrary.  For the sake of being able to make easy comparisons we should choose our value of $\ell_b$ to match the choice made in some other study.  As described earlier, the most directly comparable studies to this one are \cite{pap:chetverikov_dunkel_phase_behaviour_Morse}, in which $\ell_b = q_c$, and \cite{pap:Lepri_etal_1d_LennardJones}, in which $\ell_b = 1.5 q_c$.  As previously explained, given the experience of chain fracture studies \cite{pap:oliveira_bond_stability_criterion}, where fracture is not observed until surprisingly large values of bond length, we think there is some reason to prefer the larger value.  So we will follow \cite{pap:Lepri_etal_1d_LennardJones} and define a bond as broken if the bond length exceeds $\ell_b = 1.5 q_c$, where for the Lennard-Jones chain $q_c$ is given by (\ref{eq:qc}).  In the present section we will use nothing more than this simple geometric definition of chain breakage.  In the next section we will demonstrate a different definition which allows us to obtain additional information about the chain state.  However, the simple geometric definition presented in this section will allow us to define some terms which will be useful for any definition of chain breakage.

The probability distribution governing the $N$ bond lengths is just the position part of the total state probability distribution

\begin{equation}
\label{eq:rho_sys}
\rho_{sys} (\vec{q}) = \frac{1}{{\Phi_0}^N} \exp{\left [ -\beta \sum_{i=1}^N \left (V_{LJ} (q_i) + Pq_i \right) \right ]} \, ,
\end{equation}

\noindent which separates into $N$ independent factors, $\rho_{sys}(\vec{q}) = \prod_i \rho_i (q_i)$, where

\begin{equation}
\label{eq:single_bond_distfunc}
\rho_i (q_i) = \frac{1}{\Phi_0} \exp{\left [ -\beta (V_{LJ} (q_i) + Pq_i) \right ]} \, ,
\end{equation}

\noindent is the probability distribution for an individual bond length, $q_i$.

However, this is the probability distribution for bond lengths in an ensemble of systems.  Any specific member of the ensemble is a chain of $N$ particles with some specific length, $L$.  Since $L$ is fixed for any member of the ensemble, the joint probability distribution function for the bond lengths in an individual chain is

\begin{equation}
\rho_{\textrm{one chain}} (\vec{q}) = \frac{1}{{\Phi_0}^N} \exp{\left [ -\beta \sum_{i=1}^N \left (V_{LJ} (q_i) + Pq_i \right) \right ]} \delta \left( L - \sum_{j=1} ^N q_j \right ) \, ,
\end{equation}

\noindent The $\delta$-function prevents this from separating into individual integrals and so the bond lengths are not, strictly, independent of each other.  However, if the chain is sufficiently long it is still approximately true that the probability distribution for an individual bond length is given by (\ref{eq:single_bond_distfunc}).  If we define some breaking distance $\ell_b$ at which we say that a bond is broken then the probability that a given bond is broken is, therefore

\begin{equation}
\label{eq:Pdefn}
\mathcal{P} \equiv \mathrm{Prob}(q_i > \ell_b) \simeq \frac{1}{\Phi_0} \int_{\ell_b}^{\infty} \exp{\left [ -\beta (V_{LJ} (q_i) + Pq_i) \right ]} dq_i \, .
\end{equation}

For some specific chain at any time let the number of breaks in the chain be $n_{breaks}$.  Note that in periodic boundary conditions the number of clusters is equal to the number of breaks, except for the case of $n_{breaks} = 0$.  We can view $n_{breaks}$ as a random variable.  Since each bond is either broken with probability $\mathcal{P}$ or unbroken we can view the broken/unbroken state of each bond as a Bernoulli variable in the approximation that the probabilities of bond breakages are independent.  So $n_{breaks}$ will follow a binomial distribution

\begin{equation}
\mathrm{Prob}(n_{breaks} = n) = {{N}\choose{n}} \mathcal{P}^n \mathcal{Q}^{N-n} \, ,
\end{equation}

\noindent where $\mathcal{Q} = 1 - \mathcal{P}$.  So the average number of breaks in the chain is just $\langle n_{breaks} = N\mathcal{P} \rangle$.  In situations where one is concerned with chains of arbitrary length it is probably more useful to know the expected cluster length, $\langle K \rangle$.  The average cluster length in a chain of length , $N$, at any given time (averaged over the chain at fixed time) is simply $K_N = N/n_{breaks}$.  So we can find the expected cluster length (averaged over time) by

\begin{equation}
\langle K_N \rangle = \sum_{n=1}^N \frac{N}{n} \mathrm{Prob}(n_{breaks} = n) \, .
\end{equation}

For long chains it is inconvenient to carry out this sum.  However, for sufficiently long chains we can approximate the binomial distribution by a Gaussian distribution in the usual way

\begin{equation}
\label{eq:ave_K}
\langle K_N \rangle \simeq \frac{1}{\sqrt{2 \pi N\mathcal{PQ}}} \int_1^\infty \frac{N}{x} \exp{\left [ - \frac{(x-N\mathcal{P})^2}{2N\mathcal{PQ}} \right ]} dx \, 
\end{equation}

\noindent and in the long chain limit the distribution of $n_{breaks}$ approaches a $\delta$-function distribution so

\begin{equation}
\label{eq:lim_K}
\langle K \rangle \equiv \lim_{N \rightarrow \infty} \langle K_N \rangle = \frac{1}{\mathcal{P}} \, .
\end{equation}

\noindent Note that, while $n_{breaks}$ follows an approximately Gaussian distribution, the distribution of $K_N$ is considerably skewed with a very long right tail.  A consequence of this is that for chains of lengths that are commonly examined in numerical studies ($N \sim 10^4$ or $10^5$) this $N \rightarrow \infty$ limit is not well approached.  Even for chains of many thousands of particles the difference between (\ref{eq:ave_K}) and (\ref{eq:lim_K}) can be very large, especially if $\mathcal{P}$ is small.  However, we are most interested in chains in the thermodynamic limit.  Our interest in the thermodynamic limit stems from the fact that theories (such as mode-coupling theories \cite{pap:mine_PRE72} and fluctuating hydrodynamics \cite{pap:spohn_nonlinear_fluct_hyd}) currently being used to explain anomalous transport in one-dimensional chains are only valid in the thermodynamic limit - or at least on some length scale where there is a clear separation of scales between the scale on which local thermodynamic equilibrium is established and the scale on which transport is being examined.  Hence, from this point forward we will take, $\langle K \rangle$, defined by (\ref{eq:lim_K}), as the definition of the expected chain length.  Researchers investigating phenomena in shorter chains should use $\langle K_N \langle$, not $\langle K \rangle$.

In practice this means that if we compare values of $\mathcal{P}$ obtained from theory and simulation we find good agreement.  But if we tabulate $\langle K_N \rangle$ obtained directly from simulations it often compares poorly with theory.  In interpreting simulation results we will always calculate $\mathcal{P}$ from the simulation and then invert it to obtain a simulation value of $\langle K \rangle$.  Effectively, we are using finite chain length simulations to estimate the infinite chain limit, $\langle K \rangle$.  So long as sufficiently long chain lengths and sufficiently long simulation times are used to estimate $\mathcal{P}$, the agreement between theory and simulation is very good, as seen in Figure~\ref{fig:Kvsb_th_sim}.

We show $\langle K \rangle$ as a function of $\beta$ predicted using (\ref{eq:lim_K}) and using two different values of $\ell_b$ in Figure~\ref{fig:Kvsb_th_sim}.  As expected \cite{pap:chetverikov_dunkel_phase_behaviour_Morse}, changing $\ell_b$ makes a quantitative, but no qualitative difference to the predictions.  The figure also shows the result of using the theory developed in the next section and a comparison with simulations.

Much of the analysis in this section is applicable to any loosely bound chain, such as a Morse chain.  In particular, the only modification needed to study a different chain is to replace $V_{LJ}$ in (\ref{eq:rho_sys}) through (\ref{eq:Pdefn}) with the interparticle potential of the system of interest.  However, it should be pointed out that the checking of the results involves knowledge of the system thermodynamics as presented in section~\ref{sec:LJthermo}.  It is beyond the scope of the present paper to carry out this additional analysis.  However, one would expect that the results would be qualitatively similar to those that we obtain for the Lennard-Jones chain.  The Morse chain has been studied in \cite{pap:chetverikov_dunkel_phase_behaviour_Morse,pap:chetverikov_etal_thermo_phase_trans_Morse}, though with a rather different focus than the study carried out here.

\section{Clusters Defined by Unbound Particle Pairs}

The previous section details a way of finding probabilities of broken bonds and expected cluster sizes which is based on a bond breakage criterion similar to that used in \cite{pap:chetverikov_dunkel_phase_behaviour_Morse,pap:Lepri_etal_1d_LennardJones}.  This criterion is ``purely geometric'' in that the decision of whether a bond is broken or not is made purely on the basis of the bond length, $q_i$.  However, as pointed out in \cite{pap:chetverikov_dunkel_phase_behaviour_Morse} it may be desirable to have a criterion which includes information on bond energy.  This allows one to distinguish between a larger number of ``phases'' such as being able to distinguish between a high density gas in which only occasional bonds are broken but all clusters are unbound as opposed to a high density, but fragmented liquid.  Ultimately, one might wish to combine several criteria so as to be able to distinguish among a larger number of phases in the phase diagram.  As we will see, inclusion of bond energy makes very little difference to the prediction of $\langle K \rangle$.  However, the additional information we can get from finding $\langle K \rangle$ using bond energy as part of the criterion for bond breakage does prove to be useful for distinguishing among additional phases.

Accordingly, we will obtain expressions using an energy-based criterion for whether a bond is broken.  Let us start by defining the energy of a bond as

\begin{equation}
\label{eq:bond_energy}
E_i = \frac{1}{2}(E_{K,i} + E_{K,i+1}) + V_{LJ}(q_i) \, .
\end{equation}

\noindent That is, the energy of the $i^{th}$ bond is defined to be the potential energy of that bond plus half of the kinetic energies, $E_{K,i}$ and $E_{K, i+1}$, of the particles connected to that bond.  As a starting point let us find the probability, $\mathrm{Prob}(E_i > 0)$, that the $i^{th}$ pair of particles is ``unbound''.  If we know the probability density of the potential energy, $W(V)$, then we may write the desired probability as

\begin{equation}
\label{eq:prob_unbound}
\mathrm{Prob}(E_i > 0) \equiv \Pi = \int_{-A/4}^\infty \mathrm{Prob}(E_i >0 | V) W(V) dV \, ,
\end{equation}

\noindent where $\mathrm{Prob}(E_i > 0 | V)$ is the conditional probability that $E_i > 0$ given that $V_{LJ} = V$.  This conditional probability has the form

\begin{equation}
\label{eq:Up}
\mathrm{Prob}(E_i > 0 | V) \equiv \Upsilon = \left \{
\begin{array}{ll}
\Upsilon_{-} (\beta V), & V<0 \\
1, & V \geq 0 \, .
\end{array}
\right .
\end{equation}

\noindent where $\Upsilon_{-} = \mathrm{Prob}(E_i > 0 | V < 0) = \mathrm{Prob}(E_{K,i} + E_{K,i+1} > -2V | V < 0)$.  Thus, we can determine $\Upsilon_{-}$ from the probability density of $E_{K,i} + E_{K,i+1}$.  In the appendix, leading up to (\ref{eq:Upminus}), we show that

\begin{equation}
\label{eq:Upminus_main}
\Upsilon_{-} (\beta V) = e^{2\beta V} \, ,
\end{equation}

\noindent and that we can write $W(V)$ as

\begin{equation}
W(V) = \left \{
\begin{array}{ll}
W_+ (V) + W_{-} (V) & V<0 \\
W_{+} (V), & V \geq 0 \, .
\end{array}
\right .
\end{equation}

\noindent where $W_{+}$ and $W_{-}$ are defined in the appendix in equation (\ref{eq:Wplusminus}).  We can extract more information out of this than simply whether particles are bound or not by noting that, because of the piecewise nature of $\Upsilon (\beta V)$ and $W(V)$, (\ref{eq:prob_unbound}) can be split up according to

\begin{eqnarray}
\label{eq:Pi_defs}
\Pi & = &\int_{-A/4}^0 dV \Upsilon_{-} W_+ + \int_0^\infty dV W_+ + \int_{-A/4}^0 dV \Upsilon_{-} W_{-} \\ 
& \equiv & \Pi_{\pm} (\beta, P) + \Pi_{+} (\beta, P) + \Pi_{-} (\beta, P) \, ,
\end{eqnarray}

where we can interpret

\begin{eqnarray}
\Pi_{+} & = & \mathrm{Prob} (E_i > 0 \, \textrm{and} \, q_i \leq q_0) \, ,\\
\Pi_{\pm} & = & \mathrm{Prob} (E_i > 0 \, \textrm{and} \, q_0 < q_i < q_{min}) \, , \\
\label{eq:end_Pi_defs}
\Pi_{-} & = & \mathrm{Prob} (E_i > 0 \, \textrm{and} \, q_i \geq q_{min}) \, .
\end{eqnarray}

Furthermore, if we once again define a bond length, $\ell_b > q_{min}$, as a breaking length then we can define a mixed geometric and energetic criterion for bond breakage as that $q_i > \ell_b$ and $E_i > 0$  (the particles are separated by more than $\ell_b$ and they are unbound).  Thus, we can define a probability of bond breakage, $\Pi_b$, similarly to $\Pi_{-}$ according to

\begin{equation}
\label{eq:Pi_b}
\Pi_b = \int_{V_{LJ} (\ell_b)}^0 dV \Upsilon_{-} W_{-} \, .
\end{equation}

\begin{figure}
\includegraphics[width=\textwidth]{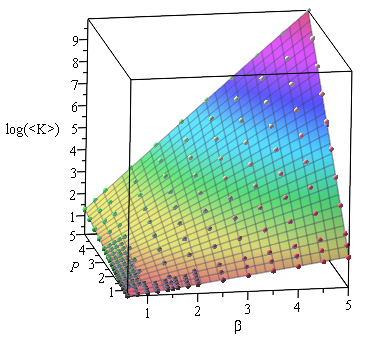}
\caption{\label{fig:KvsbP}Expected cluster length, $\langle K \rangle$, as a function of temperature and pressure using combined geometric and energetic critera for identifying breaks (probability of break $\equiv \Pi_b$).  Points show values of $\langle K \rangle$ calculated from theory.  The surface is a guide to the eye and is a Lowess-smoothed fit to the calculated points.}
\end{figure}

\noindent We can then find the expected cluster length, $\langle K \rangle$, using (\ref{eq:lim_K}) but replacing $\mathcal{P}$ defined by (\ref{eq:Pdefn}) with $\Pi_b$ defined by (\ref{eq:Pi_b}).  Evaluation of $\langle K \rangle$ at any given $(\beta, P)$ still involves numerical evaluation of integrals, but it is considerably less numerically intensive than carrying out molecular dynamics simulations to find cluster sizes or probabilities of bond breakage as has been done previously for Lennard-Jones chains in \cite{pap:Lepri_etal_1d_LennardJones} and for Morse chains in \cite{pap:chetverikov_dunkel_phase_behaviour_Morse}.  Figure~\ref{fig:KvsbP} shows $\langle K \rangle$ calculated from (\ref{eq:Pi_b}) over a range of $\beta$ and $P$.  To a good approximation, at a fixed $P$ the expected cluster length increases exponentially with $\beta$ and at fixed $\beta$ it increases exponentially with $P$.  This is consistent with the finding in \cite{pap:Lepri_etal_1d_LennardJones} that the fraction of broken bonds is $f \sim \exp{-\beta A/8}$ and makes sense from basic considerations of Boltzmann statistics.

\begin{figure}
\includegraphics[width=\textwidth]{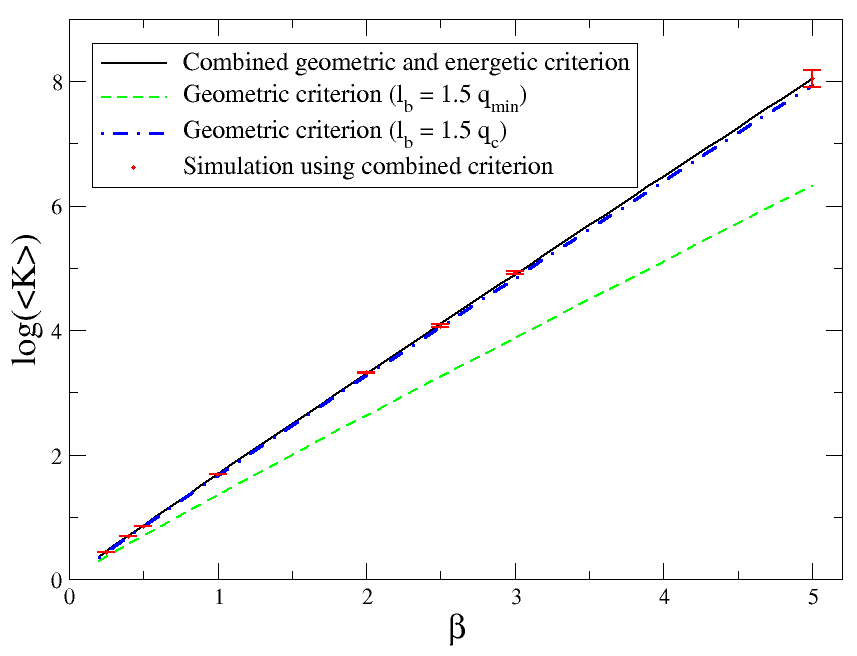}
\caption{\label{fig:Kvsb_th_sim}Expected cluster length, $\langle K \rangle$, as a function of temperature for $P=4$ using several different critera for identifying breaks.  The solid line is obtained from theory using the combined geometric and energetic criterion (probability of break $\equiv \Pi_b$).  The dashed line is obtained from theory using the geometric criterion (probability of break $\equiv \mathcal{P}$) with $\ell_b = 1.5 q_{min}$ and the dot dashed line is similar but using $\ell_b = 1.5 q_c$.  Points with error bars show values of $\langle K \rangle$ obtained from simulation using the combined criterion.  Each point is an average obtained from 10 simulation runs.  Error bars are one standard error.  All simulations used $N=2^{11}$ particles and ran for $t_{end} = 2^8$ time units using a time step of $\Delta t = 2^{-6}$ time units, except runs at $k_B/\beta = 0.2$ which used $N = 2^{20}$ particles and at $k_B/\beta = 0.3333$ which used $N = 2^{15}$ particles.  Error bars for simulation values indicate the standard error in the simulation results.}
\end{figure}

To validate the theory presented above we carry out molecular dynamics simulations.  Each simulation run is initialized into an equilibrium state generated from a constant temperature and pressure ensemble.  Particle momentums are generated from an appropriate Gaussian distribution using a Box-Muller algorithm \cite{book:num_recipes}.  Particle positions are generated from the distribution given by (\ref{eq:single_bond_distfunc}) using the rejection method \cite{book:num_recipes}.  We then compute the center of mass velocity and subtract it from every particle velocity so that the simulation ``observes'' the chain from the center of mass frame.  The system is then integrated using a 4$^{th}$ order symplectic integrator \cite{pap:LNG_JStatPhys132}.  During the runs statistics on number of broken bonds via various definitions of broken bonds are collected at regular intervals.  The total number of particle pairs that are unbound is also tracked, along with the numbers that are unbound and have bond lengths $q_i < q_0$, $q_0 \leq q_i \leq q_{min}$ and $q_i > q_{min}$.  Figure~\ref{fig:Kvsb_th_sim} shows comparison of $\langle K \rangle$ obtained from theory with simulations.  The theory and simulation both used the combined criterion for defining broken bonds.  Agreement between the theory and simulation is extremely good.  For comparison, the figure also shows theoretical predictions using the purely geometric criterion for defining broken bonds.  Simulations were used to validate these as well (not shown) and the agreement is similar to that observed for the combined criterion.  As expected, the inclusion of the energetic criterion makes little difference to the predictions of $\langle K \rangle$, and choosing different values of $\ell_b$ makes only a small difference.

\section{A Phase Diagram for the Lennard-Jones Chain}

Recall that the Lennard-Jones chain does not undergo true phase transitions \cite{pap:Lepri_etal_1d_LennardJones}.  However, in different parts of the thermodynamic $\beta P$-space we expect to find regions with distinct configurational and dynamical qualities.  In consideration of a finite chain of $N$ particles it would be reasonable to define it as being in a ``liquid-like phase'' whenever $\langle K \rangle (\beta, P) > N$.  Similarly, for a chain of any length it might be reasonable to refer to the region where $\langle K \rangle (\beta, P) \simeq 1$ as a ``gas-like phase''.  An intermediate regime, where clusters of size $K>1$ predominate, but the chain is unlikely to consist of a single cluster has been called ``liquid-like'' by others \cite{pap:chetverikov_dunkel_phase_behaviour_Morse} but we prefer to call it a ``droplet phase'' since it is analogous to water on a hydrophobic surface breaking up into droplets.  Finally, it is possible to have a phase in which $\langle K \rangle (\beta, P) \gg 1$ but most particle pairs are unbound.  This might be called a ``high density gas phase''.  These definitions are obviously not the only definitions one could come up with.  Given the lack of well defined phase transitions any choice of phase definitions is arbitrary and will be driven by the purpose behind the definitions.  We are defining these phases for illustrative purposes to show how the above theory can be used to distinguish among distinct regions of $\beta P$-space.  In the Discussion and Conclusions we will show some preliminary results examining transport in this system which demonstrate the potential usefulness of even arbitrary phase definitions like the ones adopted here.

Notice that most of the above definitions depend on the number of particles, $N$, in the chain.  An infinite chain in equilibrium at non-zero temperature is always broken into clusters, though it is possible for those clusters to be arbitrarily large.  For an infinite chain, however, it is possible to specify on any given length scale, $\lambda$, what phase it is in.  So, for example, if the temperature is low and the particle density is $\rho_n$ then on a length scale $\lambda < \langle K \rangle /\rho_n $ the system can be thought of as liquid-like whereas on longer length scales it is better thought of as being in the droplet phase.

We will define phases first of all in terms of the size of $\Pi = \Pi_{+} + \Pi_{\pm} + \Pi_{-}$ where $\Pi$, $\Pi_+$, $\Pi_-$, and $\Pi_{\pm}$ are defined in (\ref{eq:Pi_defs}) through (\ref{eq:end_Pi_defs}).  In other words, we are interested in what fraction of the particles are unbound.  If $\Pi$ is small (most particle pairs are bound) we will further subdivide between liquid and droplet phases according to $\langle K \rangle$.  If $\Pi$ is large we will subdivide between gas and high density gas phases using $\langle K \rangle$.  We define $\langle K \rangle$ using (\ref{eq:Pi_b}) as the probability of a break (mixed geometric and energetic definition of broken bonds).  We will choose the following definitions of phases, emphasizing that these are arbitrary definitions and are being used for illustrative purposes:

\begin{itemize}

\item ``Liquid-like'': $\Pi < 0.5$, $\langle K \rangle > 2^{10}$, (most particle pairs are bound, large clusters).

\item ``Droplet'': $\Pi < 0.5$, $\langle K \rangle \leq 2^{10}$, (most particle pairs are bound, small to moderate clusters).

\item ``Gas-like'': $\Pi \geq 0.5$, $ \langle K \rangle \leq 4$ (most particle pairs are unbound, tiny clusters).

\item ``High Density Gas'': $\Pi \geq 0.5$, $\langle K \rangle > 4$ (most particle pairs are unbound, moderate or large clusters).

\end{itemize}

\begin{figure}
\includegraphics[width=\textwidth]{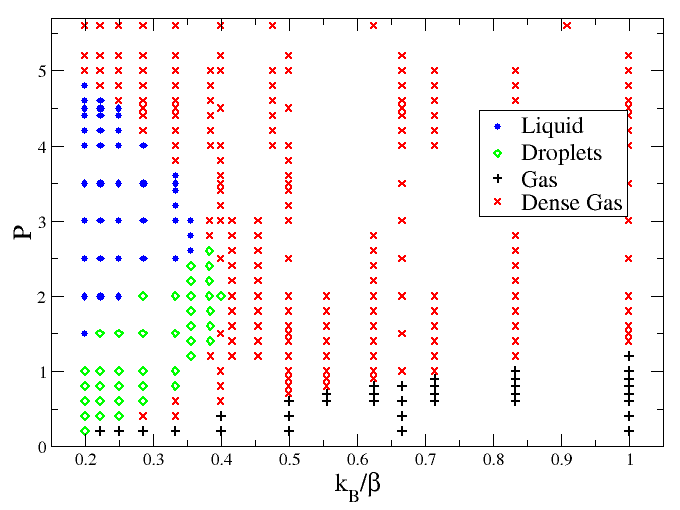}
\caption{\label{fig:phase}A possible phase diagram for the Lennard-Jones chain, based on criteria given in the text.  Solid circles indicate the liquid-like phase, unfilled diamonds indicate the droplets phase, plus signs indicate the gas-like phase, and ``X'' indicates the high density gas phase.}
\end{figure}

\noindent In particular, the choice of boundary values ($0.5$, $4$ and $2^{10}$) for $\Pi$ and $\langle K \rangle$ is completely arbitrary.  Different choices of these values could produce considerably different looking phase diagrams.  The phase diagram obtained using the above definitions is shown in Figure~\ref{fig:phase}.  We will speculate about possibly more physical criteria for distinguishing between meaningfully different regions of $\beta P$-space in the conclusions.

\section{Discussion and Conclusions}

We have shown a method for determining the phase behaviour of a one-dimensional chain.  In that the method relies only on equilibrium statistical mechanics it is quite simple.  A previous determination of a phase diagram \cite{pap:chetverikov_dunkel_phase_behaviour_Morse} for Morse chains was done via molecular dynamics simulations.  The phase diagram obtained there was qualitatively somewhat similar to the one obtained here for the Lennard-Jones chain, despite somewhat different definitions of the phases.  However, determining the phase diagram numerically as was done in \cite{pap:chetverikov_dunkel_phase_behaviour_Morse} is extremely numerically intensive and so that effort was restricted to small system sizes.  This, in turn, raises questions about the degree to which the conclusions of \cite{pap:chetverikov_dunkel_phase_behaviour_Morse} can be extended to large systems.  In contrast, the methods demonstrated in this paper are analytical.  Some computation is still required since the calculations involve integrals which must be evaluated numerically.  Nevertheless, the method is several orders of magnitude faster than numerical methods and a phase diagram can be generated on a desktop computer in a matter of minutes or hours, with the amount of time required obviously depending on how finely the $\beta P$-space is to be sampled in the diagram.

A central result of this work is the ability to determine what fraction of bonds in the chain are broken, or equivalently the expected cluster size, $\langle K \rangle$.  This was also previously done numerically \cite{pap:Lepri_etal_1d_LennardJones} and the results found here are consistent with previous reports.  Exact comparison is difficult since in \cite{pap:Lepri_etal_1d_LennardJones} the dependence of $\langle K \rangle$ with $\beta$ was found holding system density constant whereas we hold pressure constant.  However, in agreement with those earlier results we see an exponential dependence which is expected since fragmentation is expected to be a thermally activated process.

As in previous work on chains in equilibrium there is no clear definition for what constitutes a break in the chain.  Based on findings from non-equilibrium chain fracture studies \cite{pap:oliveira_bond_stability_criterion} we have chosen to use a length longer than the critical length $q_c$, in agreement with \cite{pap:Lepri_etal_1d_LennardJones}.  However, this is an extremely ad-hoc definition.  One approach might be to use explicit $\ell_b (\beta, P)$ functions obtained from non-equilibrium chain fracture studies instead of simply choosing a fixed value for $\ell_b$.  This has the virtue of being less arbitrary, at the cost of the theory being only slightly more complicated if one knows the function $\ell_b(\beta,P)$.  While it has been shown \cite{pap:oliveira_bond_stability_criterion} that $\ell_b$ depends on $\beta$ and $P$ a functional form has not been found.

Another approach for coming up with less arbitrary definitions of phases in these studies would be to use other methods to characterize the system state.  Previous studies \cite{pap:chetverikov_dunkel_phase_behaviour_Morse,pap:Lepri_etal_1d_LennardJones} have looked at the dynamical structure factor, $S(q,\omega)$.  The work so far is rather phenomenological but several conclusions can already be drawn, more or less along expected lines.  At high density and low temperatures $S(q,\omega)$ shows a single sharp phonon peak.  As the temperature increases, $S(q,\omega)$ becomes more noisy in the vicinity of some transition temperature \cite{pap:chetverikov_dunkel_phase_behaviour_Morse}.  At lower densities the low temperature $S(q,\omega)$ shows multiple peaks, likely because the chain is fragmented and each cluster behaves approximately as an isolated chain.  Also, as temperature increases a zero-frequency component appears in $S(q,\omega)$ similar to the Rayleigh peak observed in real liquids \cite{pap:Lepri_etal_1d_LennardJones}.  It might be possible to relate the appearance of these phenomenological changes to the behaviour of $\langle K \rangle$.  For example, the appearance of additional peaks at low temperatures as the density is decreased may correspond with the crossover from $\langle K \rangle > N$ to $\langle K \rangle < N$.  More speculatively, perhaps the relative areas under the Rayleigh and Brillouin peaks might be related to $\langle K \rangle / N$.

\begin{figure}
\includegraphics[width=\textwidth]{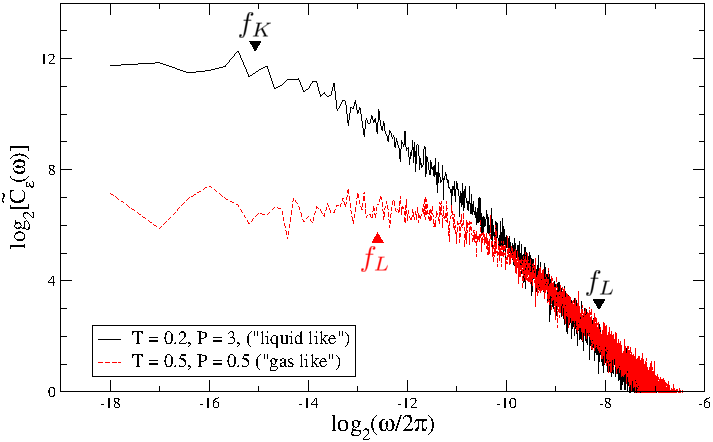}
\caption{\label{fig:powspecs}Energy current power spectra for Lennard-Jones chains under two sets of thermodynamic parameters.  Black solid curve: T = 0.2, P = 3 (indicated as liquid like in Figure~\ref{fig:phase}), red dashed curve: T = 0.5, P = 0.5 (gas like in Figure~\ref{fig:phase}).  Each curve is an average over 20 sets of initial conditions using $N = 2^{13}$ particles run for $2^{18}$ simulated time units.  Frequencies $\omega_K$ and $\omega_L$ are indicated for the black curve.  $\omega_L$ is indicated for the red curve but $\omega_K$ is not because it appears far off-scale to the right.}
\end{figure}

While the method presented here works very well for predicting expected cluster sizes, because this is entirely done using equilibrium methods it provides no information on the lifetimes of clusters.  Most attempts to understand fragmentation rates have relied on a transition state theory due to Kramers.  Early attempts to use this theory \cite{pap:oliveira_trans-state_frac_nuc} led to predictions of fragmentation rates which differed from observed rates in simulations by several orders of magnitude.  However, more recent attempts \cite{pap:sain_etal_rupture_extended_object} have resolved this by considering collective motions in obtaining an attempt frequency.  Note that this is for unbroken chains under tension.  In an equilibrium chain in a ``droplet'' phase we must have a population of clusters of different lengths.  It would be interesting to track the lifetimes of clusters of different lengths and see how well Kramers theory accounts for it.

The interest of this research group in the phase behaviour of Lennard-Jones chains stems from the study of transport phenomena.  Compared to the extensive work which has been done on thermal transport in tightly-bound oscillator chains such as the famous FPU chain \cite{pap:Lepri_review,pap:Dhar2008} there has been comparatively little work done on thermal transport in loosely bound chains such as the Lennard-Jones chain.  An early paper \cite{pap:Mareschal_Amellal_PRA37} gives rather inconclusive results and the problem seems to have been largely ignored subsequently.  Indeed, the phase behaviour of loosely bound chains almost certainly has implications for transport which cannot be ignored and this may be why these systems have proven more difficult to study.  Under conditions (low temperature, high pressure) in which $\langle K \rangle$ is large, the chain may exhibit transport phenoma similar to what is seen in the FPU chain.  After all, the FPU chain was originally conceived as nothing more than a Taylor approximation to systems governed by interactions similar to Lennard-Jones interactions.  In those systems the emerging picture is that one must consider coupling between sound modes in the system and the heat transport modes \cite{pap:LNG_JStatPhys132,pap:spohn_nonlinear_fluct_hyd}.  For a finite chain, as long as $\langle K \rangle \gg N$, simple pictures of sound transport in the system might apply.  However, if the chain is fragmented so that $\langle K \rangle < N$, sound of wavelengths smaller than the typical cluster length will be transmitted very poorly over large distances because waves cannot ``jump'' from one cluster to another.  Wavelengths on much longer length scales might be well transmitted but the primary mechanism for this would be collisions between clusters, not the coupled oscillations of individual particles.  Thus, all transport in a fragmented chain will likely have an implicit heirarchy of length scales which may couple very poorly with each other.  This is totally at odds with the more common picture of transport in one-dimensional chains.

Some evidence of the above can be seen in some preliminary numerical results from our study of transport in this system.  We hope to examine this in more detail in a future paper.  We have followed practices described in \cite{pap:mine_PRE72,pap:LNG_JStatPhys132} to numerically obtain energy current power spectra for Lennard-Jones chains in periodic boundary conditions.  These are shown in Figure~\ref{fig:powspecs}.  The figure shows spectra for two representative choices of pressure and temperature - one well within the ``liquid like'' part of Figure~\ref{fig:phase} and the other in the ``gas like'' part.  We have computed the length per particle, $\ell = L/N$, and speed of sound, $c$, for each of these parameter choices.  This allows us to find two frequencies of interest

\begin{eqnarray}
\omega_L & = & \frac{c}{N\ell} \, , \\
\omega_K & = & \frac{c}{K\ell} \, ,
\end{eqnarray}

\noindent where $\omega_L$ can be thought of as the frequency with which a long-wavelength signal propagates all the way around the chain (alternatively one can think of this as the frequency of the fundamental mode of the chain).  In contrast, $\omega_K$ is the frequency which which a long-wavelength signal travels a distance equal to the expected cluster length.  Both power spectra exhibit a low frequency saturation to a constant value.  If this saturation was caused by finite-size effects then we would expect the saturation of both to occur near $\omega_L$.  We see that in the case of the liquid like chain this is clearly not the case.  Rather, the saturation occurs near $\omega_K$.  This is what we would expect based on the physical picture described in the previous paragraph.  It is also further support for the arguments put forward in \cite{pap:mine_finite_size} regarding the validity of examining transport at frequencies lower than the system's fundamental mode frequency.  We have examined several other sets of parameters well within the liquid like part of the phase diagram and this pattern seems to hold quite generally for them.  Additionally, running simulations with different numbers of particles does not change the frequency at which the saturation occurs, which is further evidence that it is due to the cluster length, not a finite-size effect.  Note that the cluster length for this chain is much longer than the chain itself.  The saturation at $\omega_K$ seems to indicate that a disturbance can travel around the chain many times before a break ``opens up in its way'' at which point it reflects and/or damps.

In contrast, the spectrum for the gas like chain shows a saturation closer to $\omega_L$. ($\omega_K$ for this chain is not visible because it is well off-scale to the right in the graph.)  However, not much should be read into this.  Running the simulation with different chain lengths also does not change the frequency at which this saturation occurs.  Additionally, choosing different values of temperature and pressure within the gas and high density gas region produces saturations above or below $\omega_L$ with no pattern which is yet clear.  We certainly do not understand what is happening in the gas like chains, though it is clear that very low frequency contributions to the heat transport are much smaller than those in the liquid like chain.  This difference speaks to the usefulness of knowing about the phase behaviour of these chains, despite the arbitrariness of the definitions of the phases.  Based on simulations so far it is unclear whether a lower frequency regime exists, as speculated above, dominated by collisions between clusters.  The momentum current power spectra (not shown) also show interesting features, including peaks in the vicinity of $\omega_L$ for liquid like chains which are very broadened or absent in gas like chains.  This may be similar to the broadening of Brillouin lines in the dynamical structure factor of Lennard-Jones chains seen at higher temperatures in \cite{pap:Lepri_etal_1d_LennardJones}.

It has recently been claimed that the Lennard-Jones chain has a finite thermal conductivity and it has been speculated that this is a result of chain fragmentation \cite{pap:Gendelman_Savin_NormalHeatConductivityInChainsCapableOfDissociation}.  Based on the saturation of the energy current power spectrum at low frequencies in our results, especially in the gas like phase where we do not understand the reason for the saturation, it is slightly tempting to conclude that the Lennard-Jones chain might have a finite thermal conductivity.  We caution against this conclusion, mainly light of the possibility of a low frequency regime dominated by cluster-cluster collisions and other advective contributions.  Until advective contributions have been accounted for we believe that it is too early to draw conclusions about whether the Lennard-Jones chain displays regular or anomalous transport.

% Expand discussion slightly - tempting but likely wrong to conclude that Fourier's Law is obeyed.

\appendix

\section{Probability of a Pair of Particles Being Unbound}

The probability density of the $i^{th}$ momentum is just a Boltzmann distribution

\begin{equation}
f(p) = \sqrt{\frac{\beta}{2 \pi m}} \exp{\left [ \frac{-\beta {p}^2}{2m} \right ]} \, .
\end{equation}

\noindent We can, thus, find the probability density of the kinetic energy, $E_K$, using the standard theory for transformations of probability densities, where we must be careful to note that $E_K(p)$ is not one-to-one and so we must partition the domain of $E_K(p)$ into subsets that are one-to-one and construct the probability density $w (E_K)$ by summing over these partitions.  This yields the well known result

\begin{equation}
w(E_K) = \sqrt{\frac{\beta}{\pi}}\frac{e^{-\beta E_K}}{\sqrt{E_K}} \, .
\end{equation}

\noindent If, $E_{K,i}$ and $E_{K,i+1}$ are statistically independent then their joint probability distribution is just

\begin{equation}
w_{i,i+1}(E_{K,i} , E_{K,i+1}) = w_i(E_{K,i}) w_{i+1}(E_{K,i+1}) = \frac{\beta}{\pi} \frac{e^{-\beta (E_{K,i} + E_{K,i+1})}}{\sqrt{E_{K,i} E_{K,i+1}}} \, .
\end{equation}

We can find $\Upsilon_{-}$ simply by integrating $w_{i,i+1}$ over all of $E_{K,i} E_{K,i+1}$-space such that $E_{K,i} + E_{K,i+1} > - 2V$ where we note that in the calculation of $\Upsilon_{-}$, by definition $V < 0$.  Naively, based on Boltzmann statistics we expect $\Upsilon_{-} \sim f(\beta, V) \exp{[-\beta g(V)]}$, for some functions $f(\beta, V)$, $g(V)$, where dimensional considerations dictate that $g(V)$ must be linear.  It will be convenient to define $\alpha = -2\beta V > 0$.  Let us denote the region of $E_{K,i} E_{K,i+1}$-space where $E_{K,i} + E_{K,i+1} > \alpha/\beta$ as $\mathcal{R}$.  We can partition $\mathcal{R} = \Gamma_1 \cup \Gamma_2 \cup \Gamma_3$ according to

\begin{eqnarray}
\Upsilon_{-} & = & \mathrm{Prob}(\Gamma_1) + \mathrm{Prob}(\Gamma_2) + \mathrm{Prob}(\Gamma_3) \\
& = & \frac{\beta}{\pi} \left \{ \int_{\alpha/\beta}^\infty dK_2 \int_0^\infty dE_{K,1} w_{12} + \int_0^{\alpha/\beta} dE_{K,2} \int_{\alpha/\beta}^\infty dE_{K,1} w_{12} \right .\\
& & \left . + \int_0^{\alpha/\beta} dE_{K,1} \int_{\alpha/\beta-E_{K,1}}^\alpha dE_{K,2} w_{12} \right \} \, .
\end{eqnarray}

\noindent The integrals for $\mathrm{Prob}(\Gamma_1)$ and $\mathrm{Prob}(\Gamma_2)$ are easily evaluated, giving

\begin{eqnarray}
\mathrm{Prob}(\Gamma_1) & = & \erfc (\sqrt{\alpha}) \, , \\
\mathrm{Prob}(\Gamma_2) & = & \erfc (\sqrt{\alpha}) \erf (\sqrt{\alpha}) \, .
\end{eqnarray}

\noindent where $\erf(\cdot)$ is the error function and $\erfc(\cdot)$ is the complimentary error function.  Evaluating $\mathrm{Prob}(\Gamma_3)$ is somewhat more difficult.  Carrying out the integral with respect to $E_{K,2}$ we obtain

\begin{eqnarray}
\mathrm{Prob} (\Gamma_3) & = & \frac{\sqrt{\beta}}{\pi} \int_0^\alpha dx \frac{e^{-x}}{\sqrt{x}} \left ( \erf (\sqrt{\alpha}) - \erf (\sqrt{\alpha - x}) \right ) \\
& = & \left [ \erf (\sqrt{\alpha}) \right ]^2 - \frac{1}{\sqrt{\pi}} I(\alpha) \, ,
\end{eqnarray}

\noindent where $I (\alpha)$ can be determined by expanding the error function in a series \cite{book:abramowitz_stegun}

\begin{eqnarray}
\label{eq:I_alpha}
I (\alpha) & = & \int_0^\alpha dx \frac{e^{-x}}{\sqrt{x}} \erf (\sqrt{\alpha - x}) \\
& = & \frac{2}{\sqrt{\pi}} \int_0^\alpha dx \frac{e^{-x}}{\sqrt{x}} e^{-(\alpha -x)} \sum_{j=0}^\infty \frac{2^n}{(2j+1)!!}(\alpha-x)^j \\
& = & \frac{2}{\pi} e^{-\alpha} \sum_{j = 0}^\infty \psi_j (\alpha) \, 
\end{eqnarray}

\noindent where $\psi_j (\alpha)$ is the integral

\begin{equation}
\psi_j (\alpha) = \int_0^\alpha du \sqrt{\frac{u}{\alpha-u}} \left [ \frac{2^j}{(2j+1)!!} u^j \right ] = \frac{\pi}{2(j+1)!} \alpha^{j+1} \, ,
\end{equation}

\noindent so that we can recognize the sum over $j$ as a Taylor expansion of an exponential function $\sum_j \psi_j = (\pi/2) (e^\alpha - 1)$.  Inserting this back into (\ref{eq:I_alpha}) we finally obtain

\begin{equation}
I(\alpha) = \sqrt{\pi} (1 - e^{-\alpha}) \, .
\end{equation}

We can now write $\Upsilon_{-}$ as

\begin{eqnarray}
\Upsilon_{-} & = & \erfc (\sqrt{\alpha}) + \erfc (\sqrt{\alpha}) \erf (\sqrt{\alpha}) + \left [ \erf (\sqrt{\alpha})\right ]^2 + e^{-\alpha} - 1 \\
\label{eq:Upminus}
& = & e^{-\alpha} = e^{2\beta V}
\end{eqnarray}

\noindent where we have used $\erf (x) + \erfc (x) = 1$ two times in the final step.  This equation is duplicated above as (\ref{eq:Upminus_main}).  This is in line with our naive expectation, though perhaps simpler than expected.  The surprisingly simple final form of this strongly suggests that there must be a much easier way to obtain it.

% \begin{figure}
% \includegraphics[width=\textwidth]{R_partition_diagram}
% \caption{\label{fig:Rparts}Partitioning the region $\mathcal{R}$.}
% \end{figure}

Returning to the problem of evaluating (\ref{eq:prob_unbound}), we must now find the probability density, $W(V)$.  Again, a bond length $q$ follows a known distribution and $V=V_{LJ}(q)$, so this is an exercise in elementary probability theory, where we must again note that $V_{LJ} (q)$ is not one-to-one and so we will have to partition into subdomains.  Carrying this out we obtain

\begin{eqnarray}
\label{eq:W}
W(V) & = & \rho_i [r_+ (V)] \left | \frac{d r_+}{dV} \right | + \rho_i [r_{-} (V)] \left | \frac{dr_{-}}{dV} \right | \\
\label{eq:Wplusminus}
& \equiv & W_+ (V) + W_{-} (V)
\end{eqnarray}

\noindent where $r_{\pm}$ are the two inversions of $V_{LJ} (r)$

\begin{equation}
r_{\pm} = \left [ \frac{2}{1 \pm \sqrt{1+4V/A}} \right ]^{1/6} q_0 \, ,
\end{equation}

\noindent and

\begin{equation}
\rho_i \left [ r_\pm (V) \right ] = \frac{1}{Y} \exp \left [ -\beta (V + P r_{\pm} (V)) \right ] \, .
\end{equation} 

\noindent The form (\ref{eq:W}) for $W(V)$ is only valid for $V < 0$, which is mapped from $q > q_0$.  For $V \geq 0$ ($q \leq q_0$) we simply have $W(V) = W_{+} (V)$. 

Combining (\ref{eq:prob_unbound}), (\ref{eq:Up}), (\ref{eq:Upminus}), and (\ref{eq:W}) will yield the desired probability that a pair of particles is unbound.  Numerically integrating (\ref{eq:prob_unbound}) must be done with care; at high temperatures and/or low pressures the integrand has an extremely tall and narrow peak just to the left of $V=0$ caused by the large number of long bond lengths in the system under those conditions.

\begin{acknowledgments}
This work was supported by a grant from Cape Breton University.  Simulations were carried out on ACENet, a high perfomance computing consortium operated by several Canadian universities.
\end{acknowledgments}

%\bibliography{refs}

\begin{thebibliography}{50}%
\makeatletter
\providecommand \@ifxundefined [1]{%
 \@ifx{#1\undefined}
}%
\providecommand \@ifnum [1]{%
 \ifnum #1\expandafter \@firstoftwo
 \else \expandafter \@secondoftwo
 \fi
}%
\providecommand \@ifx [1]{%
 \ifx #1\expandafter \@firstoftwo
 \else \expandafter \@secondoftwo
 \fi
}%
\providecommand \natexlab [1]{#1}%
\providecommand \enquote  [1]{``#1''}%
\providecommand \bibnamefont  [1]{#1}%
\providecommand \bibfnamefont [1]{#1}%
\providecommand \citenamefont [1]{#1}%
\providecommand \href@noop [0]{\@secondoftwo}%
\providecommand \href [0]{\begingroup \@sanitize@url \@href}%
\providecommand \@href[1]{\@@startlink{#1}\@@href}%
\providecommand \@@href[1]{\endgroup#1\@@endlink}%
\providecommand \@sanitize@url [0]{\catcode `\\12\catcode `\$12\catcode
  `\&12\catcode `\#12\catcode `\^12\catcode `\_12\catcode `\%12\relax}%
\providecommand \@@startlink[1]{}%
\providecommand \@@endlink[0]{}%
\providecommand \url  [0]{\begingroup\@sanitize@url \@url }%
\providecommand \@url [1]{\endgroup\@href {#1}{\urlprefix }}%
\providecommand \urlprefix  [0]{URL }%
\providecommand \Eprint [0]{\href }%
\providecommand \doibase [0]{http://dx.doi.org/}%
\providecommand \selectlanguage [0]{\@gobble}%
\providecommand \bibinfo  [0]{\@secondoftwo}%
\providecommand \bibfield  [0]{\@secondoftwo}%
\providecommand \translation [1]{[#1]}%
\providecommand \BibitemOpen [0]{}%
\providecommand \bibitemStop [0]{}%
\providecommand \bibitemNoStop [0]{.\EOS\space}%
\providecommand \EOS [0]{\spacefactor3000\relax}%
\providecommand \BibitemShut  [1]{\csname bibitem#1\endcsname}%
\let\auto@bib@innerbib\@empty
%</preamble>
\bibitem [{\citenamefont {Ford}(1992)}]{pap:Ford_review}%
  \BibitemOpen
  \bibfield  {author} {\bibinfo {author} {\bibfnamefont {J.}~\bibnamefont
  {Ford}},\ }\href@noop {} {\bibfield  {journal} {\bibinfo  {journal} {Phys.
  Rep.}\ }\textbf {\bibinfo {volume} {213}},\ \bibinfo {pages} {271} (\bibinfo
  {year} {1992})}\BibitemShut {NoStop}%
\bibitem [{\citenamefont
  {Lepri}(2016)}]{book:Lepri_ThermalTransportLowDimensions}%
  \BibitemOpen
  \bibinfo {editor} {\bibfnamefont {S.}~\bibnamefont {Lepri}},\ ed.,\
  \href@noop {} {\emph {\bibinfo {title} {Thermal transport in low dimensions:
  from statistical physics to nanoscale heat transfer}}},\ Vol.\ \bibinfo
  {volume} {921}\ (\bibinfo  {publisher} {Springer},\ \bibinfo {year}
  {2016})\BibitemShut {NoStop}%
\bibitem [{\citenamefont {Hahn}\ \emph {et~al.}(1996)\citenamefont {Hahn},
  \citenamefont {Karger},\ and\ \citenamefont
  {Kukla}}]{pap:hahn_etal_single-file_diffusion_observation}%
  \BibitemOpen
  \bibfield  {author} {\bibinfo {author} {\bibfnamefont {K.}~\bibnamefont
  {Hahn}}, \bibinfo {author} {\bibfnamefont {J.}~\bibnamefont {Karger}}, \ and\
  \bibinfo {author} {\bibfnamefont {V.}~\bibnamefont {Kukla}},\ }\href@noop {}
  {\bibfield  {journal} {\bibinfo  {journal} {Phys. Rev. Lett.}\ }\textbf
  {\bibinfo {volume} {76}},\ \bibinfo {pages} {2762} (\bibinfo {year}
  {1996})}\BibitemShut {NoStop}%
\bibitem [{\citenamefont {Metzler}\ and\ \citenamefont
  {Klafter}(2000)}]{pap:metzler_klafter_random_walk_guide_anomalous_diffusion}%
  \BibitemOpen
  \bibfield  {author} {\bibinfo {author} {\bibfnamefont {R.}~\bibnamefont
  {Metzler}}\ and\ \bibinfo {author} {\bibfnamefont {J.}~\bibnamefont
  {Klafter}},\ }\href@noop {} {\bibfield  {journal} {\bibinfo  {journal} {Phys.
  Rep.}\ }\textbf {\bibinfo {volume} {339}},\ \bibinfo {pages} {1} (\bibinfo
  {year} {2000})}\BibitemShut {NoStop}%
\bibitem [{\citenamefont {Lee-Dadswell}\ \emph {et~al.}(2008)\citenamefont
  {Lee-Dadswell}, \citenamefont {Nickel},\ and\ \citenamefont
  {Gray}}]{pap:LNG_JStatPhys132}%
  \BibitemOpen
  \bibfield  {author} {\bibinfo {author} {\bibfnamefont {G.~R.}\ \bibnamefont
  {Lee-Dadswell}}, \bibinfo {author} {\bibfnamefont {B.~G.}\ \bibnamefont
  {Nickel}}, \ and\ \bibinfo {author} {\bibfnamefont {C.~G.}\ \bibnamefont
  {Gray}},\ }\href@noop {} {\bibfield  {journal} {\bibinfo  {journal} {J. Stat.
  Phys.}\ }\textbf {\bibinfo {volume} {132}},\ \bibinfo {pages} {1} (\bibinfo
  {year} {2008})}\BibitemShut {NoStop}%
\bibitem [{\citenamefont {Dhar}(2008)}]{pap:Dhar2008}%
  \BibitemOpen
  \bibfield  {author} {\bibinfo {author} {\bibfnamefont {A.}~\bibnamefont
  {Dhar}},\ }\href@noop {} {\bibfield  {journal} {\bibinfo  {journal} {Adv.
  Phys.}\ }\textbf {\bibinfo {volume} {57}},\ \bibinfo {pages} {457} (\bibinfo
  {year} {2008})}\BibitemShut {NoStop}%
\bibitem [{\citenamefont {Chiu}\ \emph {et~al.}(2005)\citenamefont {Chiu},
  \citenamefont {Deshpande}, \citenamefont {Postma}, \citenamefont {Lau},
  \citenamefont {Miko}, \citenamefont {Forro},\ and\ \citenamefont
  {Bockrath}}]{pap:chiu_etal_ballistic_phonon_thermal_transport_carbon_nanotube}%
  \BibitemOpen
  \bibfield  {author} {\bibinfo {author} {\bibfnamefont {H.-Y.}\ \bibnamefont
  {Chiu}}, \bibinfo {author} {\bibfnamefont {V.~V.}\ \bibnamefont {Deshpande}},
  \bibinfo {author} {\bibfnamefont {H.~W.~C.}\ \bibnamefont {Postma}}, \bibinfo
  {author} {\bibfnamefont {C.~N.}\ \bibnamefont {Lau}}, \bibinfo {author}
  {\bibfnamefont {C.}~\bibnamefont {Miko}}, \bibinfo {author} {\bibfnamefont
  {L.}~\bibnamefont {Forro}}, \ and\ \bibinfo {author} {\bibfnamefont
  {M.}~\bibnamefont {Bockrath}},\ }\href@noop {} {\bibfield  {journal}
  {\bibinfo  {journal} {Phys. Rev. Lett.}\ }\textbf {\bibinfo {volume} {95}},\
  \bibinfo {pages} {226101} (\bibinfo {year} {2005})}\BibitemShut {NoStop}%
\bibitem [{\citenamefont {Bonetto}\ \emph {et~al.}(2000)\citenamefont
  {Bonetto}, \citenamefont {Lebowitz},\ and\ \citenamefont
  {Rey-Bellet}}]{pap:Bonetto_Lebowitz_FourierLaw}%
  \BibitemOpen
  \bibfield  {author} {\bibinfo {author} {\bibfnamefont {F.}~\bibnamefont
  {Bonetto}}, \bibinfo {author} {\bibfnamefont {J.~L.}\ \bibnamefont
  {Lebowitz}}, \ and\ \bibinfo {author} {\bibfnamefont {L.}~\bibnamefont
  {Rey-Bellet}},\ }in\ \href@noop {} {\emph {\bibinfo {booktitle} {Mathematical
  Physics 2000}}},\ \bibinfo {editor} {edited by\ \bibinfo {editor}
  {\bibfnamefont {A.}~\bibnamefont {Fokas}}, \bibinfo {editor} {\bibfnamefont
  {A.}~\bibnamefont {Grigoryan}}, \bibinfo {editor} {\bibfnamefont
  {T.}~\bibnamefont {Kibble}}, \ and\ \bibinfo {editor} {\bibfnamefont
  {B.}~\bibnamefont {Zegarlinski}}}\ (\bibinfo  {publisher} {Imperial College
  Press},\ \bibinfo {year} {2000})\ pp.\ \bibinfo {pages}
  {128--150}\BibitemShut {NoStop}%
\bibitem [{\citenamefont {Zabusky}\ and\ \citenamefont
  {Kruskal}(1965)}]{pap:Zabusky_Kruskal_PRL15}%
  \BibitemOpen
  \bibfield  {author} {\bibinfo {author} {\bibfnamefont {N.~J.}\ \bibnamefont
  {Zabusky}}\ and\ \bibinfo {author} {\bibfnamefont {M.~D.}\ \bibnamefont
  {Kruskal}},\ }\href@noop {} {\bibfield  {journal} {\bibinfo  {journal} {Phys.
  Rev. Lett.}\ }\textbf {\bibinfo {volume} {15}},\ \bibinfo {pages} {240}
  (\bibinfo {year} {1965})}\BibitemShut {NoStop}%
\bibitem [{\citenamefont {Emery}\ and\ \citenamefont
  {Axe}(1978)}]{pap:EmeryAxe_PRL40}%
  \BibitemOpen
  \bibfield  {author} {\bibinfo {author} {\bibfnamefont {V.~J.}\ \bibnamefont
  {Emery}}\ and\ \bibinfo {author} {\bibfnamefont {J.~D.}\ \bibnamefont
  {Axe}},\ }\href@noop {} {\bibfield  {journal} {\bibinfo  {journal} {Phys.
  Rev. Lett.}\ }\textbf {\bibinfo {volume} {40}},\ \bibinfo {pages} {1507}
  (\bibinfo {year} {1978})}\BibitemShut {NoStop}%
\bibitem [{\citenamefont {Yoshida}\ \emph {et~al.}(1981)\citenamefont
  {Yoshida}, \citenamefont {Shobu},\ and\ \citenamefont
  {Mori}}]{pap:YoshidaShobuMori}%
  \BibitemOpen
  \bibfield  {author} {\bibinfo {author} {\bibfnamefont {T.}~\bibnamefont
  {Yoshida}}, \bibinfo {author} {\bibfnamefont {K.}~\bibnamefont {Shobu}}, \
  and\ \bibinfo {author} {\bibfnamefont {H.}~\bibnamefont {Mori}},\ }\href@noop
  {} {\bibfield  {journal} {\bibinfo  {journal} {Prog. Theoret. Phys.}\
  }\textbf {\bibinfo {volume} {66}},\ \bibinfo {pages} {759} (\bibinfo {year}
  {1981})}\BibitemShut {NoStop}%
\bibitem [{\citenamefont {Valiullin}\ and\ \citenamefont
  {K{\:a}rger}(2008)}]{Valiullin_Karger_inKlages_AnomalousMolecularDynamicsConfinedSpaces}%
  \BibitemOpen
  \bibfield  {author} {\bibinfo {author} {\bibfnamefont {R.}~\bibnamefont
  {Valiullin}}\ and\ \bibinfo {author} {\bibfnamefont {J.}~\bibnamefont
  {K{\:a}rger}},\ }in\ \href@noop {} {\emph {\bibinfo {booktitle} {Anomalous
  transport: foundations and applications}}},\ \bibinfo {editor} {edited by\
  \bibinfo {editor} {\bibfnamefont {R.}~\bibnamefont {Klages}}, \bibinfo
  {editor} {\bibfnamefont {G.}~\bibnamefont {Radons}}, \ and\ \bibinfo {editor}
  {\bibfnamefont {I.~M.}\ \bibnamefont {Sokolov}}}\ (\bibinfo  {publisher}
  {Wiley-VCH},\ \bibinfo {year} {2008})\ Chap.~\bibinfo {chapter} {18}, pp.\
  \bibinfo {pages} {519--544}\BibitemShut {NoStop}%
\bibitem [{\citenamefont {Cahill}\ \emph {et~al.}(2003)\citenamefont {Cahill},
  \citenamefont {Ford}, \citenamefont {Goodson}, \citenamefont {Mahan},
  \citenamefont {Majumdar}, \citenamefont {Maris}, \citenamefont {Merlin},\
  and\ \citenamefont {Phillpot}}]{pap:cahill_etal_nanoscale_thermal_transport}%
  \BibitemOpen
  \bibfield  {author} {\bibinfo {author} {\bibfnamefont {D.~G.}\ \bibnamefont
  {Cahill}}, \bibinfo {author} {\bibfnamefont {W.~K.}\ \bibnamefont {Ford}},
  \bibinfo {author} {\bibfnamefont {K.~E.}\ \bibnamefont {Goodson}}, \bibinfo
  {author} {\bibfnamefont {G.~D.}\ \bibnamefont {Mahan}}, \bibinfo {author}
  {\bibfnamefont {A.}~\bibnamefont {Majumdar}}, \bibinfo {author}
  {\bibfnamefont {H.~J.}\ \bibnamefont {Maris}}, \bibinfo {author}
  {\bibfnamefont {R.}~\bibnamefont {Merlin}}, \ and\ \bibinfo {author}
  {\bibfnamefont {S.~R.}\ \bibnamefont {Phillpot}},\ }\href@noop {} {\bibfield
  {journal} {\bibinfo  {journal} {J. Appl. Phys.}\ }\textbf {\bibinfo {volume}
  {93}},\ \bibinfo {pages} {793} (\bibinfo {year} {2003})}\BibitemShut
  {NoStop}%
\bibitem [{\citenamefont {Liu}\ \emph {et~al.}(2012)\citenamefont {Liu},
  \citenamefont {Xu}, \citenamefont {Xie}, \citenamefont {Zhang},\ and\
  \citenamefont
  {Li}}]{pap:Liu_etal_AnomalousHeatConductionAndAnomalousDiffusionInLowDimensionalNanoscaleSystems}%
  \BibitemOpen
  \bibfield  {author} {\bibinfo {author} {\bibfnamefont {S.}~\bibnamefont
  {Liu}}, \bibinfo {author} {\bibfnamefont {X.~F.}\ \bibnamefont {Xu}},
  \bibinfo {author} {\bibfnamefont {R.~G.}\ \bibnamefont {Xie}}, \bibinfo
  {author} {\bibfnamefont {G.}~\bibnamefont {Zhang}}, \ and\ \bibinfo {author}
  {\bibfnamefont {B.~W.}\ \bibnamefont {Li}},\ }\href@noop {} {\bibfield
  {journal} {\bibinfo  {journal} {Eur. Phys. J. B}\ } (\bibinfo {year}
  {2012})}\BibitemShut {NoStop}%
\bibitem [{\citenamefont {Meier}\ \emph {et~al.}(2014)\citenamefont {Meier},
  \citenamefont {Menges}, \citenamefont {Nirmalraj}, \citenamefont {Holscher},
  \citenamefont {Riel},\ and\ \citenamefont
  {Gotsmann}}]{pap:Meier_etal_LengthDependentTransportInPolymerChains_Experiment_PRL113}%
  \BibitemOpen
  \bibfield  {author} {\bibinfo {author} {\bibfnamefont {T.}~\bibnamefont
  {Meier}}, \bibinfo {author} {\bibfnamefont {F.}~\bibnamefont {Menges}},
  \bibinfo {author} {\bibfnamefont {P.}~\bibnamefont {Nirmalraj}}, \bibinfo
  {author} {\bibfnamefont {H.}~\bibnamefont {Holscher}}, \bibinfo {author}
  {\bibfnamefont {H.}~\bibnamefont {Riel}}, \ and\ \bibinfo {author}
  {\bibfnamefont {B.}~\bibnamefont {Gotsmann}},\ }\href@noop {} {\bibfield
  {journal} {\bibinfo  {journal} {Phys. Rev. Lett.}\ }\textbf {\bibinfo
  {volume} {113}},\ \bibinfo {pages} {060801} (\bibinfo {year}
  {2014})}\BibitemShut {NoStop}%
\bibitem [{\citenamefont {Lepri}\ \emph {et~al.}(2003)\citenamefont {Lepri},
  \citenamefont {Livi},\ and\ \citenamefont {Politi}}]{pap:Lepri_review}%
  \BibitemOpen
  \bibfield  {author} {\bibinfo {author} {\bibfnamefont {S.}~\bibnamefont
  {Lepri}}, \bibinfo {author} {\bibfnamefont {R.}~\bibnamefont {Livi}}, \ and\
  \bibinfo {author} {\bibfnamefont {A.}~\bibnamefont {Politi}},\ }\href@noop {}
  {\bibfield  {journal} {\bibinfo  {journal} {Phys. Rep.}\ }\textbf {\bibinfo
  {volume} {377}},\ \bibinfo {pages} {1} (\bibinfo {year} {2003})}\BibitemShut
  {NoStop}%
\bibitem [{\citenamefont
  {Lee-Dadswell}(2015{\natexlab{a}})}]{pap:mine_universality_classes}%
  \BibitemOpen
  \bibfield  {author} {\bibinfo {author} {\bibfnamefont {G.~R.}\ \bibnamefont
  {Lee-Dadswell}},\ }\href@noop {} {\bibfield  {journal} {\bibinfo  {journal}
  {Phys. Rev. E}\ }\textbf {\bibinfo {volume} {91}},\ \bibinfo {pages} {032102}
  (\bibinfo {year} {2015}{\natexlab{a}})}\BibitemShut {NoStop}%
\bibitem [{\citenamefont {Pr\:{a}hofer}\ and\ \citenamefont
  {Spohn}(2004)}]{pap:Prahofer_Spohn_ExactScalingFunctionsForOne-DimensionalStationaryKPZ}%
  \BibitemOpen
  \bibfield  {author} {\bibinfo {author} {\bibfnamefont {M.}~\bibnamefont
  {Pr\:{a}hofer}}\ and\ \bibinfo {author} {\bibfnamefont {H.}~\bibnamefont
  {Spohn}},\ }\href@noop {} {\bibfield  {journal} {\bibinfo  {journal} {Journal
  of Statistical Physics}\ }\textbf {\bibinfo {volume} {115}},\ \bibinfo
  {pages} {255–279} (\bibinfo {year} {2004})}\BibitemShut {NoStop}%
\bibitem [{\citenamefont {Casati}\ and\ \citenamefont
  {Prosen}(2003)}]{pap:Casati_Prosen_AnomalousConductionHardPointGas}%
  \BibitemOpen
  \bibfield  {author} {\bibinfo {author} {\bibfnamefont {G.}~\bibnamefont
  {Casati}}\ and\ \bibinfo {author} {\bibfnamefont {T.}~\bibnamefont
  {Prosen}},\ }\href@noop {} {\bibfield  {journal} {\bibinfo  {journal} {Phys.
  Rev. E}\ }\textbf {\bibinfo {volume} {67}},\ \bibinfo {pages} {015203(R)}
  (\bibinfo {year} {2003})}\BibitemShut {NoStop}%
\bibitem [{\citenamefont {Casati}\ \emph {et~al.}(1984)\citenamefont {Casati},
  \citenamefont {Ford}, \citenamefont {Vivaldi},\ and\ \citenamefont
  {Visscher}}]{pap:CasatiFordetal}%
  \BibitemOpen
  \bibfield  {author} {\bibinfo {author} {\bibfnamefont {G.}~\bibnamefont
  {Casati}}, \bibinfo {author} {\bibfnamefont {J.}~\bibnamefont {Ford}},
  \bibinfo {author} {\bibfnamefont {F.}~\bibnamefont {Vivaldi}}, \ and\
  \bibinfo {author} {\bibfnamefont {W.~M.}\ \bibnamefont {Visscher}},\
  }\href@noop {} {\bibfield  {journal} {\bibinfo  {journal} {Phys. Rev. Lett.}\
  }\textbf {\bibinfo {volume} {52}},\ \bibinfo {pages} {1861} (\bibinfo {year}
  {1984})}\BibitemShut {NoStop}%
\bibitem [{\citenamefont {Hu}\ \emph {et~al.}(1998)\citenamefont {Hu},
  \citenamefont {Li},\ and\ \citenamefont
  {Zhao}}]{pap:HuLiZhao_frenkel_kontorova}%
  \BibitemOpen
  \bibfield  {author} {\bibinfo {author} {\bibfnamefont {B.}~\bibnamefont
  {Hu}}, \bibinfo {author} {\bibfnamefont {B.}~\bibnamefont {Li}}, \ and\
  \bibinfo {author} {\bibfnamefont {H.}~\bibnamefont {Zhao}},\ }\href@noop {}
  {\bibfield  {journal} {\bibinfo  {journal} {Phys. Rev. E}\ }\textbf {\bibinfo
  {volume} {57}},\ \bibinfo {pages} {2992} (\bibinfo {year}
  {1998})}\BibitemShut {NoStop}%
\bibitem [{\citenamefont {Giardin\'a}\ \emph {et~al.}(2000)\citenamefont
  {Giardin\'a}, \citenamefont {Livi}, \citenamefont {Politi},\ and\
  \citenamefont {Vassalli}}]{GiardinaLivietal}%
  \BibitemOpen
  \bibfield  {author} {\bibinfo {author} {\bibfnamefont {C.}~\bibnamefont
  {Giardin\'a}}, \bibinfo {author} {\bibfnamefont {R.}~\bibnamefont {Livi}},
  \bibinfo {author} {\bibfnamefont {A.}~\bibnamefont {Politi}}, \ and\ \bibinfo
  {author} {\bibfnamefont {M.}~\bibnamefont {Vassalli}},\ }\href@noop {}
  {\bibfield  {journal} {\bibinfo  {journal} {Phys. Rev. Lett.}\ }\textbf
  {\bibinfo {volume} {84}},\ \bibinfo {pages} {2144} (\bibinfo {year}
  {2000})}\BibitemShut {NoStop}%
\bibitem [{\citenamefont {Lee-Dadswell}\ \emph {et~al.}(2010)\citenamefont
  {Lee-Dadswell}, \citenamefont {Turner}, \citenamefont {Ettinger},\ and\
  \citenamefont {Moy}}]{pap:mine_mcdl}%
  \BibitemOpen
  \bibfield  {author} {\bibinfo {author} {\bibfnamefont {G.~R.}\ \bibnamefont
  {Lee-Dadswell}}, \bibinfo {author} {\bibfnamefont {E.}~\bibnamefont
  {Turner}}, \bibinfo {author} {\bibfnamefont {J.}~\bibnamefont {Ettinger}}, \
  and\ \bibinfo {author} {\bibfnamefont {M.}~\bibnamefont {Moy}},\ }\href@noop
  {} {\bibfield  {journal} {\bibinfo  {journal} {Phys. Rev. E}\ }\textbf
  {\bibinfo {volume} {82}},\ \bibinfo {pages} {061118} (\bibinfo {year}
  {2010})}\BibitemShut {NoStop}%
\bibitem [{\citenamefont {Bishop}(1982)}]{pap:bishop_collective_modes_LJ}%
  \BibitemOpen
  \bibfield  {author} {\bibinfo {author} {\bibfnamefont {M.}~\bibnamefont
  {Bishop}},\ }\href@noop {} {\bibfield  {journal} {\bibinfo  {journal} {J.
  Stat. Phys.}\ }\textbf {\bibinfo {volume} {29}},\ \bibinfo {pages} {623}
  (\bibinfo {year} {1982})}\BibitemShut {NoStop}%
\bibitem [{\citenamefont {Mareschal}\ and\ \citenamefont
  {Amellal}(1988)}]{pap:Mareschal_Amellal_PRA37}%
  \BibitemOpen
  \bibfield  {author} {\bibinfo {author} {\bibfnamefont {M.}~\bibnamefont
  {Mareschal}}\ and\ \bibinfo {author} {\bibfnamefont {A.}~\bibnamefont
  {Amellal}},\ }\href@noop {} {\bibfield  {journal} {\bibinfo  {journal} {Phys.
  Rev. A}\ }\textbf {\bibinfo {volume} {37}},\ \bibinfo {pages} {2189}
  (\bibinfo {year} {1988})}\BibitemShut {NoStop}%
\bibitem [{\citenamefont {Li}\ \emph {et~al.}(2015)\citenamefont {Li},
  \citenamefont {Liu}, \citenamefont {Li}, \citenamefont {Hanggi},\ and\
  \citenamefont {Li}}]{pap:Li_etal_ConundrumOfAnomalousVsNormalHeatTransport}%
  \BibitemOpen
  \bibfield  {author} {\bibinfo {author} {\bibfnamefont {Y.}~\bibnamefont
  {Li}}, \bibinfo {author} {\bibfnamefont {S.}~\bibnamefont {Liu}}, \bibinfo
  {author} {\bibfnamefont {N.}~\bibnamefont {Li}}, \bibinfo {author}
  {\bibfnamefont {P.}~\bibnamefont {Hanggi}}, \ and\ \bibinfo {author}
  {\bibfnamefont {B.}~\bibnamefont {Li}},\ }\href@noop {} {\bibfield  {journal}
  {\bibinfo  {journal} {New J. Phys,}\ }\textbf {\bibinfo {volume} {17}},\
  \bibinfo {pages} {043064} (\bibinfo {year} {2015})}\BibitemShut {NoStop}%
\bibitem [{\citenamefont {Gendelman}\ and\ \citenamefont
  {Savin}(2014)}]{pap:Gendelman_Savin_NormalHeatConductivityInChainsCapableOfDissociation}%
  \BibitemOpen
  \bibfield  {author} {\bibinfo {author} {\bibfnamefont {O.~V.}\ \bibnamefont
  {Gendelman}}\ and\ \bibinfo {author} {\bibfnamefont {A.~V.}\ \bibnamefont
  {Savin}},\ }\href@noop {} {\bibfield  {journal} {\bibinfo  {journal}
  {Europhys. Lett.}\ }\textbf {\bibinfo {volume} {106}},\ \bibinfo {pages}
  {34004} (\bibinfo {year} {2014})}\BibitemShut {NoStop}%
\bibitem [{\citenamefont {Dunkel}\ \emph {et~al.}(2002)\citenamefont {Dunkel},
  \citenamefont {Ebeling}, \citenamefont {Erdmann},\ and\ \citenamefont
  {Makarov}}]{pap:dunkel_etal_coherent_motions_dissipative_Morse}%
  \BibitemOpen
  \bibfield  {author} {\bibinfo {author} {\bibfnamefont {J.}~\bibnamefont
  {Dunkel}}, \bibinfo {author} {\bibfnamefont {W.}~\bibnamefont {Ebeling}},
  \bibinfo {author} {\bibfnamefont {U.}~\bibnamefont {Erdmann}}, \ and\
  \bibinfo {author} {\bibfnamefont {V.~A.}\ \bibnamefont {Makarov}},\
  }\href@noop {} {\bibfield  {journal} {\bibinfo  {journal} {Int. J.
  Bifurcation Chaos}\ }\textbf {\bibinfo {volume} {12}},\ \bibinfo {pages}
  {2359} (\bibinfo {year} {2002})}\BibitemShut {NoStop}%
\bibitem [{\citenamefont
  {Pereira}(2010)}]{pap:pereira_anharmonic_crystals_as_thermal_diodes}%
  \BibitemOpen
  \bibfield  {author} {\bibinfo {author} {\bibfnamefont {E.}~\bibnamefont
  {Pereira}},\ }\href@noop {} {\bibfield  {journal} {\bibinfo  {journal} {Phys.
  Rev. E}\ }\textbf {\bibinfo {volume} {82}},\ \bibinfo {pages} {040101͑(R͒)}
  (\bibinfo {year} {2010})}\BibitemShut {NoStop}%
\bibitem [{\citenamefont {Spohn}(2014)}]{pap:spohn_nonlinear_fluct_hyd}%
  \BibitemOpen
  \bibfield  {author} {\bibinfo {author} {\bibfnamefont {H.}~\bibnamefont
  {Spohn}},\ }\href@noop {} {\bibfield  {journal} {\bibinfo  {journal} {J.
  Stat. Phys.}\ }\textbf {\bibinfo {volume} {154}},\ \bibinfo {pages} {1191}
  (\bibinfo {year} {2014})}\BibitemShut {NoStop}%
\bibitem [{\citenamefont {Delfini}\ \emph {et~al.}(2006)\citenamefont
  {Delfini}, \citenamefont {Lepri}, \citenamefont {Livi},\ and\ \citenamefont
  {Politi}}]{pap:Delfini_etal}%
  \BibitemOpen
  \bibfield  {author} {\bibinfo {author} {\bibfnamefont {L.}~\bibnamefont
  {Delfini}}, \bibinfo {author} {\bibfnamefont {S.}~\bibnamefont {Lepri}},
  \bibinfo {author} {\bibfnamefont {R.}~\bibnamefont {Livi}}, \ and\ \bibinfo
  {author} {\bibfnamefont {A.}~\bibnamefont {Politi}},\ }\href@noop {}
  {\bibfield  {journal} {\bibinfo  {journal} {Phys. Rev. E}\ }\textbf {\bibinfo
  {volume} {73}},\ \bibinfo {pages} {060201(R)} (\bibinfo {year}
  {2006})}\BibitemShut {NoStop}%
\bibitem [{\citenamefont
  {Dyson}(1969)}]{pap:dyson_existence_of_phase_transition_one-dimension}%
  \BibitemOpen
  \bibfield  {author} {\bibinfo {author} {\bibfnamefont {F.~J.}\ \bibnamefont
  {Dyson}},\ }\href@noop {} {\bibfield  {journal} {\bibinfo  {journal} {Comm.
  Math. Phys.}\ }\textbf {\bibinfo {volume} {12}},\ \bibinfo {pages} {91}
  (\bibinfo {year} {1969})}\BibitemShut {NoStop}%
\bibitem [{\citenamefont {Wang}\ and\ \citenamefont
  {Casati}(2017)}]{pap:WangCasati_1DPhaseTransition}%
  \BibitemOpen
  \bibfield  {author} {\bibinfo {author} {\bibfnamefont {J.}~\bibnamefont
  {Wang}}\ and\ \bibinfo {author} {\bibfnamefont {G.}~\bibnamefont {Casati}},\
  }\href@noop {} {\bibfield  {journal} {\bibinfo  {journal} {Phys. Rev. Lett.}\
  }\textbf {\bibinfo {volume} {118}},\ \bibinfo {pages} {040601} (\bibinfo
  {year} {2017})}\BibitemShut {NoStop}%
\bibitem [{\citenamefont {{Van
  Hove}}(1950)}]{pap:van_hove_physica_no_one-dimensional_phase_transition}%
  \BibitemOpen
  \bibfield  {author} {\bibinfo {author} {\bibfnamefont {L.}~\bibnamefont {{Van
  Hove}}},\ }\href@noop {} {\bibfield  {journal} {\bibinfo  {journal}
  {Physica}\ }\textbf {\bibinfo {volume} {16}},\ \bibinfo {pages} {137}
  (\bibinfo {year} {1950})}\BibitemShut {NoStop}%
\bibitem [{\citenamefont {Landau}\ and\ \citenamefont
  {Lifshitz}(1958)}]{LandauLifshitz_StatPhys}%
  \BibitemOpen
  \bibfield  {author} {\bibinfo {author} {\bibfnamefont {L.~D.}\ \bibnamefont
  {Landau}}\ and\ \bibinfo {author} {\bibfnamefont {E.~M.}\ \bibnamefont
  {Lifshitz}},\ }\href@noop {} {\emph {\bibinfo {title} {Statistical Physics
  {I}}}},\ \bibinfo {edition} {1st}\ ed.\ (\bibinfo  {publisher} {Pergamon
  Press},\ \bibinfo {address} {Oxford},\ \bibinfo {year} {1958})\BibitemShut
  {NoStop}%
\bibitem [{\citenamefont {Chetverikov}\ and\ \citenamefont
  {Dunkel}(2003)}]{pap:chetverikov_dunkel_phase_behaviour_Morse}%
  \BibitemOpen
  \bibfield  {author} {\bibinfo {author} {\bibfnamefont {A.}~\bibnamefont
  {Chetverikov}}\ and\ \bibinfo {author} {\bibfnamefont {J.}~\bibnamefont
  {Dunkel}},\ }\href@noop {} {\bibfield  {journal} {\bibinfo  {journal} {Eur.
  Phys. J. B}\ }\textbf {\bibinfo {volume} {35}},\ \bibinfo {pages} {239}
  (\bibinfo {year} {2003})}\BibitemShut {NoStop}%
\bibitem [{\citenamefont
  {Stillinger}(1995)}]{pap:stillinger_stat_mech_of_metastable_matter}%
  \BibitemOpen
  \bibfield  {author} {\bibinfo {author} {\bibfnamefont {F.~H.}\ \bibnamefont
  {Stillinger}},\ }\href@noop {} {\bibfield  {journal} {\bibinfo  {journal}
  {Phys. Rev. E}\ }\textbf {\bibinfo {volume} {52}},\ \bibinfo {pages} {4685}
  (\bibinfo {year} {1995})}\BibitemShut {NoStop}%
\bibitem [{\citenamefont {Lepri}\ \emph {et~al.}(2005)\citenamefont {Lepri},
  \citenamefont {Sandri},\ and\ \citenamefont
  {Politi}}]{pap:Lepri_etal_1d_LennardJones}%
  \BibitemOpen
  \bibfield  {author} {\bibinfo {author} {\bibfnamefont {S.}~\bibnamefont
  {Lepri}}, \bibinfo {author} {\bibfnamefont {P.}~\bibnamefont {Sandri}}, \
  and\ \bibinfo {author} {\bibfnamefont {A.}~\bibnamefont {Politi}},\
  }\href@noop {} {\bibfield  {journal} {\bibinfo  {journal} {Eur. Phys. J. B}\
  }\textbf {\bibinfo {volume} {47}},\ \bibinfo {pages} {549} (\bibinfo {year}
  {2005})}\BibitemShut {NoStop}%
\bibitem [{\citenamefont {Chetverikov}\ \emph {et~al.}(2005)\citenamefont
  {Chetverikov}, \citenamefont {Ebeling},\ and\ \citenamefont
  {Velarde}}]{pap:chetverikov_etal_thermo_phase_trans_Morse}%
  \BibitemOpen
  \bibfield  {author} {\bibinfo {author} {\bibfnamefont {A.~P.}\ \bibnamefont
  {Chetverikov}}, \bibinfo {author} {\bibfnamefont {W.}~\bibnamefont
  {Ebeling}}, \ and\ \bibinfo {author} {\bibfnamefont {M.~G.}\ \bibnamefont
  {Velarde}},\ }\href@noop {} {\bibfield  {journal} {\bibinfo  {journal} {Eur.
  Phys. J. B}\ }\textbf {\bibinfo {volume} {44}},\ \bibinfo {pages} {509}
  (\bibinfo {year} {2005})}\BibitemShut {NoStop}%
\bibitem [{\citenamefont {Oliveira}(1998)}]{pap:oliveira_trans-state_frac_nuc}%
  \BibitemOpen
  \bibfield  {author} {\bibinfo {author} {\bibfnamefont {F.~A.}\ \bibnamefont
  {Oliveira}},\ }\href@noop {} {\bibfield  {journal} {\bibinfo  {journal}
  {Phys. Rev. B}\ }\textbf {\bibinfo {volume} {57}},\ \bibinfo {pages} {10576}
  (\bibinfo {year} {1998})}\BibitemShut {NoStop}%
\bibitem [{\citenamefont {Welland}\ \emph {et~al.}(1992)\citenamefont
  {Welland}, \citenamefont {Shin}, \citenamefont {Allen},\ and\ \citenamefont
  {Ketterson}}]{pap:welland_etal_fracture_in_1d}%
  \BibitemOpen
  \bibfield  {author} {\bibinfo {author} {\bibfnamefont {R.~W.}\ \bibnamefont
  {Welland}}, \bibinfo {author} {\bibfnamefont {M.}~\bibnamefont {Shin}},
  \bibinfo {author} {\bibfnamefont {D.}~\bibnamefont {Allen}}, \ and\ \bibinfo
  {author} {\bibfnamefont {J.~B.}\ \bibnamefont {Ketterson}},\ }\href@noop {}
  {\bibfield  {journal} {\bibinfo  {journal} {Phys. Rev. B}\ }\textbf {\bibinfo
  {volume} {46}},\ \bibinfo {pages} {503} (\bibinfo {year} {1992})}\BibitemShut
  {NoStop}%
\bibitem [{\citenamefont {Oliveira}\ and\ \citenamefont
  {Gonzalez}(1996)}]{pap:oliveira_bond_stability_criterion}%
  \BibitemOpen
  \bibfield  {author} {\bibinfo {author} {\bibfnamefont {F.~A.}\ \bibnamefont
  {Oliveira}}\ and\ \bibinfo {author} {\bibfnamefont {J.~A.}\ \bibnamefont
  {Gonzalez}},\ }\href@noop {} {\bibfield  {journal} {\bibinfo  {journal}
  {Phys. Rev. B}\ }\textbf {\bibinfo {volume} {54}},\ \bibinfo {pages} {3954}
  (\bibinfo {year} {1996})}\BibitemShut {NoStop}%
\bibitem [{\citenamefont {Sain}\ \emph {et~al.}(2006)\citenamefont {Sain},
  \citenamefont {Dias},\ and\ \citenamefont
  {Grant}}]{pap:sain_etal_rupture_extended_object}%
  \BibitemOpen
  \bibfield  {author} {\bibinfo {author} {\bibfnamefont {A.}~\bibnamefont
  {Sain}}, \bibinfo {author} {\bibfnamefont {C.~L.}\ \bibnamefont {Dias}}, \
  and\ \bibinfo {author} {\bibfnamefont {M.}~\bibnamefont {Grant}},\
  }\href@noop {} {\bibfield  {journal} {\bibinfo  {journal} {Phys. Rev. E}\
  }\textbf {\bibinfo {volume} {74}},\ \bibinfo {pages} {046111} (\bibinfo
  {year} {2006})}\BibitemShut {NoStop}%
\bibitem [{\citenamefont {Bazhenov}\ and\ \citenamefont
  {Heyes}(1990)}]{pap:bazhenov_heyes_dynprops_transcoeffs_LJ}%
  \BibitemOpen
  \bibfield  {author} {\bibinfo {author} {\bibfnamefont {A.~M.}\ \bibnamefont
  {Bazhenov}}\ and\ \bibinfo {author} {\bibfnamefont {D.~M.}\ \bibnamefont
  {Heyes}},\ }\href@noop {} {\bibfield  {journal} {\bibinfo  {journal} {J.
  Chem. Phys.}\ }\textbf {\bibinfo {volume} {92}},\ \bibinfo {pages} {1106}
  (\bibinfo {year} {1990})}\BibitemShut {NoStop}%
\bibitem [{\citenamefont {Kim}\ and\ \citenamefont
  {Tadmor}(2012)}]{pap:kim_tadmor_analytical_free_energy_1d_chain}%
  \BibitemOpen
  \bibfield  {author} {\bibinfo {author} {\bibfnamefont {W.~K.}\ \bibnamefont
  {Kim}}\ and\ \bibinfo {author} {\bibfnamefont {E.~B.}\ \bibnamefont
  {Tadmor}},\ }\href@noop {} {\bibfield  {journal} {\bibinfo  {journal} {J.
  Stat. Phys.}\ }\textbf {\bibinfo {volume} {148}},\ \bibinfo {pages} {951}
  (\bibinfo {year} {2012})}\BibitemShut {NoStop}%
\bibitem [{\citenamefont {Oliveira}\ and\ \citenamefont
  {Taylor}(1994)}]{pap:oliveira_taylor_breaking_polymer_chains}%
  \BibitemOpen
  \bibfield  {author} {\bibinfo {author} {\bibfnamefont {F.~A.}\ \bibnamefont
  {Oliveira}}\ and\ \bibinfo {author} {\bibfnamefont {P.~L.}\ \bibnamefont
  {Taylor}},\ }\href@noop {} {\bibfield  {journal} {\bibinfo  {journal} {J.
  Chem. Phys.}\ }\textbf {\bibinfo {volume} {101}},\ \bibinfo {pages} {10118}
  (\bibinfo {year} {1994})}\BibitemShut {NoStop}%
\bibitem [{\citenamefont {Lee-Dadswell}\ \emph {et~al.}(2005)\citenamefont
  {Lee-Dadswell}, \citenamefont {Nickel},\ and\ \citenamefont
  {Gray}}]{pap:mine_PRE72}%
  \BibitemOpen
  \bibfield  {author} {\bibinfo {author} {\bibfnamefont {G.~R.}\ \bibnamefont
  {Lee-Dadswell}}, \bibinfo {author} {\bibfnamefont {B.~G.}\ \bibnamefont
  {Nickel}}, \ and\ \bibinfo {author} {\bibfnamefont {C.~G.}\ \bibnamefont
  {Gray}},\ }\href@noop {} {\bibfield  {journal} {\bibinfo  {journal} {Phys.
  Rev. E}\ }\textbf {\bibinfo {volume} {72}},\ \bibinfo {pages} {031202}
  (\bibinfo {year} {2005})}\BibitemShut {NoStop}%
\bibitem [{\citenamefont {Press}\ \emph {et~al.}(2007)\citenamefont {Press},
  \citenamefont {Teukolsky}, \citenamefont {Vetterling},\ and\ \citenamefont
  {Flannery}}]{book:num_recipes}%
  \BibitemOpen
  \bibfield  {author} {\bibinfo {author} {\bibfnamefont {W.~H.}\ \bibnamefont
  {Press}}, \bibinfo {author} {\bibfnamefont {S.~A.}\ \bibnamefont
  {Teukolsky}}, \bibinfo {author} {\bibfnamefont {W.~T.}\ \bibnamefont
  {Vetterling}}, \ and\ \bibinfo {author} {\bibfnamefont {B.~P.}\ \bibnamefont
  {Flannery}},\ }\href@noop {} {\emph {\bibinfo {title} {Numerical Recipes: The
  Art of Scientific Computing}}},\ \bibinfo {edition} {3rd}\ ed.\ (\bibinfo
  {publisher} {Cambridge University Press},\ \bibinfo {year}
  {2007})\BibitemShut {NoStop}%
\bibitem [{\citenamefont
  {Lee-Dadswell}(2015{\natexlab{b}})}]{pap:mine_finite_size}%
  \BibitemOpen
  \bibfield  {author} {\bibinfo {author} {\bibfnamefont {G.~R.}\ \bibnamefont
  {Lee-Dadswell}},\ }\href@noop {} {\bibfield  {journal} {\bibinfo  {journal}
  {Phys. Rev. E}\ }\textbf {\bibinfo {volume} {91}},\ \bibinfo {pages} {012138}
  (\bibinfo {year} {2015}{\natexlab{b}})}\BibitemShut {NoStop}%
\bibitem [{\citenamefont {Abramowitz}\ and\ \citenamefont
  {Stegun}(1964)}]{book:abramowitz_stegun}%
  \BibitemOpen
  \bibfield  {author} {\bibinfo {author} {\bibfnamefont {M.}~\bibnamefont
  {Abramowitz}}\ and\ \bibinfo {author} {\bibfnamefont {I.}~\bibnamefont
  {Stegun}},\ }\href@noop {} {\emph {\bibinfo {title} {Handbook of mathematical
  functions with formulas, graphs, and mathematical tables}}}\ (\bibinfo
  {publisher} {National Bureau of Standards},\ \bibinfo {address}
  {Gaithersburg, Maryland},\ \bibinfo {year} {1964})\BibitemShut {NoStop}%
\end{thebibliography}
%

%\begin{thebibliography}{10}

%\end{thebibliography}

\end{document}